\documentclass[pra,twocolumn,showpacs]{revtex4}

\usepackage{amsmath,amssymb,bm,graphicx,amsthm}

\newcommand{\ket}[1]{|#1\rangle}
\newcommand{\bra}[1]{\langle#1|}

\newcommand{\Tr}{\mathrm{tr}}

\newcommand{\mc}{\mathcal}
\newcommand{\SSR}{\mathrm{SSR}}
\newcommand{\tr}{\mathrm{tr}}
\newcommand{\CEPR}{\ket{\mbox{\rm E-EPR}}}
\newcommand{\VEPR}{\ket{\mbox{\rm V-EPR}}}
\newcommand{\smsqrt}[1]{\textstyle\sqrt{#1}}
\newcommand{\rhoSep}{\rho_\mathrm{sep}}
\newcommand{\rhoCoh}{\rho_\mathrm{coh}}

\newtheorem{theorem}{Theorem}
\newtheorem{definition}{Definition}
\newtheorem{corollary}{Corollary}

\begin{document}

\title{Quantum entanglement theory in the presence of superselection rules}
\author{Norbert \surname{Schuch}}
\affiliation{Max-Planck-Institut f\"ur Quantenoptik,
  Hans-Kopfermann-Str.\ 1, D--85748 Garching, Germany.}
\author{Frank \surname{Verstraete}}
\affiliation{Max-Planck-Institut f\"ur Quantenoptik,
  Hans-Kopfermann-Str.\ 1, D--85748 Garching, Germany.}
\author{J.\ Ignacio \surname{Cirac}}
\affiliation{Max-Planck-Institut f\"ur Quantenoptik,
  Hans-Kopfermann-Str.\ 1, D--85748 Garching, Germany.}

\begin{abstract}
Superselection rules severly constrain the operations which can be
implemented on a distributed quantum system. While the restriction to
local operations and classical communication gives rise to entanglement as
a nonlocal resource, particle number conservation additionally confines
the possible operations and should give rise to a new resource. In [Phys.\
Rev.\ Lett.\ \textbf{92}, 087904 (2004), \mbox{quant-ph/0310124}] 
we showed that this resource can
be quantified by a single additional number, the superselection induced
variance (SiV) without changing the concept of entanglement. In this
paper, we give the results on pure states in greater detail; additionally,
we provide a discussion of mixed state nonlocality with superselection
rules where we consider both formation and distillation.  Finally, we
demonstrate that SiV is indeed a resource, i.e., that it captures how well
a state can be used to overcome the restrictions imposed by the
superselection rule.

\end{abstract}

\pacs{03.67.-a,03.65.Ud,03.67.Mn}

\maketitle

\section{Introduction}

One of the most interesting results in quantum information theory has been
the discovery that the amount of nonlocality contained in a bipartite
quantum system can be quantified by a single number, the entropy of
entanglement (EoE). Asymptotically, multiple copies of any two states can
be converted into each other and thus into singlets provided that the
total EoE is conserved~\cite{bennett:purestate-ent}. On the other hand,
entanglement is the key resource for some of the most interesting tasks in
quantum information, as teleportation~\cite{bennett:teleportation} and
dense coding~\cite{bennett:dense_coding}.

Entanglement has its origin in the restriction to those transformations
which can be implemented by local operations and classical communication
(LOCC)~\cite{zanardi:operator-subsystems:M,barnum:subsystems:M}.  
In the same way, any additional restriction should lead to
another nonlocal quantity and thus to new effects and applications. It has
been noted by Popescu~\cite{popescu:private} that in many physical systems
of interest such a restriction is given by a superselection rule (SSR). In
this work, we will consider particle number conservation as a
superselection rule; this is motivated, e.g., by recent quantum optical
experiments on cold atomic gases. Indeed, the notion of entanglement is
affected by the additional
restrictions~\cite{verstraete:SSR-datahiding,wiseman:SSR:M}, and new
protocols arise, e.g., perfect data hiding~\cite{terhal:datahiding}
becomes possible~\cite{verstraete:SSR-datahiding}. On the other hand, it
has been shown~\cite{verstraete:SSR-datahiding,kitaev:reference_frame}
that the extra resource of a shared reference frame (i.e., a nonlocal
state) allows to overcome the restrictions imposed by the SSR (note
that~\cite{kitaev:reference_frame} also adressed non-Abelian SSR); 
conversely, private reference frames restrict the possible operations of an
eavesdropper and can thus be employed for cryptographic
tasks~\cite{bartlett:D-full}.

In~\cite{schuch:SiV-purestate}, we have shown that the nonlocality
contained in a bipartite state subject to SSR can be quantified by only
one additional number, the \emph{superselection induced variance} (SiV):
any two states can be interconverted asymptotically as long as the total
EoE \emph{and} SiV are conserved. In this paper, we prove this result in
greater detail and extend it to mixed states. We start by discussing how
the majorization criterion~\cite{nielsen:majorization} which governs the
conversion of quantum states has to be changed when SSR are present, and
show that it asymptotically converges to the conservation of EoE (as it is
the case without SSR) and SiV. We give a detailed proof of this result for
arbitrary states and show that it motivates the definition of two
different types of standard forms for SiV which carry a linear resp.\
logarithmic amount of EoE.

While there exist pure states which carry only EoE, there are no pure
states which contain solely SiV. On the other hand, it has been
demonstrated~\cite{verstraete:SSR-datahiding} that there exist separable
but nonlocal mixed states, i.e., states which have a separable
decomposition and thus do not contain EoE, but are still nonlocal as
all these decompositions violate the SSR and therefore should contain SiV.
In order to make
these statements quantitative, we extend the concepts of EoE and SiV 
to mixed states subject to SSR. One natural way to do this is to
consider the amount of pure states resources needed to create the
state~\cite{bennett:form-rec-hashing}; we show that this extension can be
done in a meaningful way and that there indeed exist states which contain
SiV but no EoE. The converse way is to ask whether it is possible
to distill pure state resources from some mixed
state~\cite{bennett:recurrence}; we provide ways to distill both EoE and
SiV, and we show that it is even possible to distill the SiV contained in
separable states.

EoE is a \emph{resource}---it
allows to overcome the LOCC restrictions by teleportation. It is
reasonable to assume that any restriction leads to a nonlocal quantity 
which in turn allows to overcome this restriction. Indeed, we give evidence
that SiV can be used as a resource which allows to overcome the
additional restrictions imposed by the SSR in a bipartite setting 
(cf.~\cite{verstraete:SSR-datahiding,schuch:SiV-purestate}). 
Therefore, we
will use two tasks: distinguishing locally undistinguishable quantum
states and teleporting states with nonconstant local particle
number~\cite{verstraete:SSR-datahiding}. We will show that not only pure
states can be used as share reference frames for these tasks, but that
there even exist separable states which together with one
ebit of entanglement allow to perfectly teleport one qubit and thus to
overcome
all restrictions. Still, we find that there is a fundamental difference
between EoE and SiV as a resource, as a finite amount of nonlocality does
not allow to perfectly overcome the restrictions which is due to the
structure of the underlying Hilbert space~\cite{schuch:SiV-purestate}.

The paper is organized as follows. In Sec.~\ref{section:definitions}, we
introduce the concept of a superselection rule and show how it 
restricts the operations which can be implemented in a bipartite setting. In
Sec.~\ref{section:pure}, we consider the conversion of pure states.
We start with the conversion of single copies, which motivates the
definition of SiV as a nonlocal monotone; then, we prove that
asymptotically all states can be converted given that both SiV and EoE are
conserved. Sec.~\ref{section:mixed} is
devoted to mixed state nonlocality. First, we discuss formation of mixed
states; beyond other results, we provide explicit formulas for the case of
qubits. Second, we give different methods for the distillation of both EoE
and SiV independently as well as simultaneously.  Finally,
Sec.~\ref{section:resource} discusses SiV as a resource; there, we
quantify how well states with SiV can be used as
shared reference frames which allow to overcome the new restrictions, and
we demonstrate that one ebit of entanglement is still sufficient for teleportation.

%%%%%%%%%%%%%%%%%%%%%%%%%%%%%%%%%%%%%%%%%%%%%%%%%%%%%%%%%%%%%%%%%%%%%%%%%
%%%%%%%%%%%%%%%%%%%%%%%%%%%%%%%%%%%%%%%%%%%%%%%%%%%%%%%%%%%%%%%%%%%%%%%%%
\section{Particle number conservation as a superselection rule
\label{section:definitions}}
%%%%%%%%%%%%%%%%%%%%%%%%%%%%%%%%%%%%%%%%%%%%%%%%%%%%%%%%%%%%%%%%%%%%%%%%%
%%%%%%%%%%%%%%%%%%%%%%%%%%%%%%%%%%%%%%%%%%%%%%%%%%%%%%%%%%%%%%%%%%%%%%%%%

In this paper, we focus on particle number conservation as a SSR, but the 
results also apply to charge and other discrete quantities.
In this case, the Hilbert space of the system $\mc
H$ can be decomposed into a direct sum $\mc H=\bigoplus_{N=0}^{\infty}\mc
H_N$ of the eigenspaces of the particle number operator $\hat N$, and 
the SSR imposes that for 
any operator $\mc O$, $[\mc O,\hat N]=0$ must hold; thus, any
operator can be written as a sum of operators $\mc O_N$ which have support
on $\mc H_N$ only,
$\mc O=\bigoplus_{N=0}^\infty\mc O_N$,
and thus 
\begin{equation}
\label{eq:OisPnOPn}
\mc O=\sum_NP_N\mc OP_N\ ,
\end{equation}
where $P_N$ projects onto $\mc H_N$.  As the same restriction holds for
the admissible density operators, all states can be converted into each
other, and no interesting new effects can be found.

Therefore, we consider SSR in a bipartite setting.  Then, we have
local particle number operators $\hat N_A$ and $\hat N_B$, and the
total particle number operator is given by
\begin{equation}
\label{eq:N_additivity}
\hat N_{AB}=\hat N_A\otimes\openone_B+\openone_A\otimes\hat N_B\ .
\end{equation}
While the admissible states have to commute with the global
particle number operator $\hat N_{AB}$, 
the local operations have to commute with the 
local particle number operators $\hat N_A$ and $\hat N_B$.
This restriction is stronger than the one given by the bipartite setting
alone and should therefore lead to a new nonlocal resource.
More precisely, the operations on subspaces with fixed
total particle number $N=N_{AB}$ are given by
\begin{equation}
\label{eq:OABNBipSSR}
\mc O^{AB}_N=
    \bigoplus_{N_A+N_B=N}
	\left(\mc O^A_{N_A}\otimes\openone^B_{N_B}\right)
\end{equation}
(and vice versa)---in addition to the restriction to products
$\mc O^A\otimes\openone$ imposed by the bipartite setting, a direct sum
structure arises from the SSR. This product vs.\ sum structure
will reappear throughout the paper and is the reason for some 
fundamental differences between EoE (arising for the product structure) and
SiV (arising from the direct sum).

The restriction to block-diagonal operations, Eq.~(\ref{eq:OisPnOPn}), can
be relaxed by adding ancilla modes with $m_0$ particles, performing a block-diagonal
unitary $U$, and measuring resp.\ tracing out the ancillas. Then, the
admissible (POVM/Kraus) operators are given by 
$\mc O=P^{\mathrm{anc}}_{m}UP^\mathrm{anc}_{m_0}$; by applying~(\ref{eq:OisPnOPn})
to $U$, 
this leads to $\mc O=\sum_NP_{N+\Delta}\mc OP_N$ (resp.\
$[\hat N,\mc O]=\Delta\mc O$, $\Delta$ might differ for each
$\mc O$): $\mc O$ can shift the particle number by some $\Delta$. (Note
that $\mc O^\dagger\mc O$ remains block-diagonal). As most results of this
paper are only affected marginally by including ancillas, we will usually
neglect them and just briefly comment on their effect as appropriate.

At the end of this section, let us introduce a few notational conventions.
Logarithms are taken to the basis $2$.
A ket $\ket N$ denotes a state with $N$ particles.  We will use this
notation even if the underlying eigenspace is degenerate, unless the
nonlocal properties under consideration depend on this degeneracy.

The restrictions imposed by the SSR on the allowed operations can be
easily overcome by defining a new computational basis 
$\ket{\hat 0}\equiv\ket{01}$, $\ket{\hat 1}\equiv\ket{10}$
in which all states have the same 
particle number~\cite{verstraete:SSR-datahiding}. 
This motivates the definition of \emph{two different}
types of maximally entangled two-qubit states, 
$$
\VEPR=\ket0_A\ket1_B+\ket1_A\ket0_B
$$ 
(a ``variance-EPR'', as there is some variance in the local particle
number), and
$$
\CEPR=\ket{01}_A\ket{10}_B+\ket{10}_A\ket{01}_B
    \equiv\ket{\hat0}_A\ket{\hat1}_B+\ket{\hat1}_A\ket{\hat0}_B
$$ 
(an ``entanglement-EPR'', which is defined within an unrestricted subspace
and only carries entanglement).
The very difference between these two states will be a central issue of the 
paper.

\section{Characterization of pure states
\label{section:pure}}

In this section, we characterize  pure states in a bipartite setting,
i.e., we determine the possible conversions by LOCC and thus quantify the
nonlocality contained in a bipartite state.  Without superselection rules,
the majorization criterion determines whether the conversion between two
bipartite pure states is
possible~\cite{nielsen:majorization,jonathan:ent_trafo}.  The conversion
of multiple copies is governed by a much simpler criterion: it has been
shown~\cite{bennett:purestate-ent} that multiple copies of any two states
can be interconverted reversibly. The conversion ratio is determined by
only one quantity which fully characterizes the nonlocal properties of a
bipartite state, the \emph{entropy of entanglement} (EoE).

As we have seen in the preceding section, in addition to the tensor
product structure induced by the bipartite setting the operators have to
obey a direct sum structure.  In this section, we show that these two
structures lead to two complementary resources: while the tensor product
again induces the majorization criterion and (asymptotically) EoE as a
nonlocal resource, the direct sum gives rise to additional restrictions on
the conversions of states and in turn leads to an own nonlocal resource.

\subsection{The single copy case}

Let us consider the following problem: given pure
bipartite states $\phi$ and $\psi$, is it possible to convert $\phi$ to
$\psi$ by LOCC? This task can be generalized naturally to a set of
outcomes $\{(p_i,\psi_i)\}$, where each outcome $\psi_i$ is obtained with
probability $p_i$. 

Let us first see how this can be solved without
SSR~\cite{nielsen:majorization}.  Therefore, let
$\bm\lambda=(\lambda_k)$ and $\bm\mu^i=(\mu_k^i)$ be the Schmidt
coefficients of $\phi$ and $\psi_i$, respectively, which completely
characterize the states up to local unitaries.
Without loss of generality, the Schmidt vectors may be taken decreasing
($\lambda_k\ge\lambda_{k+1}$) and of equal dimension (by appending zeros). 
Following~\cite{nielsen:majorization}, an LOCC strategy for the conversion
$$
\phi\longrightarrow\{(p_i,\psi_i)\}
$$
exists if and only if (iff)
$$
\bm\lambda\prec\sum_ip_i\bm\mu^i\ .
$$
Here, for two ordered vectors $\bm\lambda$ and $\bm\mu$, we say that
\emph{$\bm\lambda$ is majorized by $\bm\mu$}, $\bm\lambda\prec\bm\mu$, if
$\sum_{k=1}^d\lambda_k\le\sum_{k=1}^d\mu_k$ for all 
$1\le d<\mathrm{dim}\ \bm\lambda$, where equality holds for
$d=\mathrm{dim}\ \bm\lambda$.

As an example, consider the states
\begin{eqnarray*}
\ket\phi&=&\smsqrt{\frac12}\ket0_A\ket1_B+\smsqrt{\frac12}\ket1_A\ket0_B\
\ \mbox{and}\\
\ket\psi&=&\smsqrt{\frac13}\;\ket0_A\ket1_B+\smsqrt{\frac23}\;\ket1_A\ket0_B\ ,
\end{eqnarray*}
which have the ordered Schmidt vectors $\bm\lambda=(1/2,1/2)$ and
$\bm\mu=(2/3,1/3)$, respectively. Since $\bm\lambda\prec\bm\mu$, it is possible to
convert $\phi\rightarrow\psi$; for instance,  Alice might start with the POVM
measurement given by
$M_1=\sqrt{1/3}\ket0\bra0+\sqrt{2/3}\ket1\bra1$ and
$M_2=\sqrt{1/3}\ket0\bra0+\sqrt{2/3}\ket1\bra1$ 
which yields the two states
\begin{eqnarray*}
\ket{\psi_1}&=&\smsqrt{\frac13}\;\ket0_A\ket1_B+\smsqrt{\frac23}\;\ket1_A\ket0_B\mbox{\ \ and}\\
\ket{\psi_2}&=&\smsqrt{\frac23}\;\ket0_A\ket1_B+\smsqrt{\frac13}\;\ket1_A\ket0_B\ .
\end{eqnarray*}
with equal probabilities: $\psi_1$ is already equal to $\psi$, and
$\psi_2$ can be converted to $\psi$ by a bilateral \textsc{not} operation.

Let us now see what is different when SSR apply:
while the POVM measurement $\{M_1,M_2\}$ is compatible with the
superselection rule, the local application of \textsc{not} operations is
not; indeed, it is not possible at all to carry out $\phi\rightarrow\psi$
deterministically in the presence of SSR. In order to see this, define 
block-diagonal POVM operators $M_i=\bigoplus_n M^i_n$ on one local system.
Then, the completeness relation $\sum M_i^\dagger M_i=\openone$ yields 
$\sum_i M^{i\dagger}_n M^i_n=\openone$ for all $n$. Therefore, any POVM
operator is simply a direct sum of POVM operators acting within the 
subspaces with constant local particle number, i.e., the usual conditions
for convertibility have to hold for each subspace separately.
Particularly, this implies that for pure states the \emph{average weight}
of each subspace with constant local particle number \emph{cannot be
changed} by local operations. 

The impossibility to change the average weight of a subspace with fixed
local particle number can even be proven at a much more fundamental level.
Take multiple copies of some state $\ket\phi$ with nonconstant local
particle number, and assume  there is a way for Alice to change her local
particle number distribution on average. As the total particle
number is constant, this implies that the average particle number
distribution of Bob's system is changed the other way round.  
Therewith, Alice can change Bob's density matrix remotely
which would allow for supraluminal communication and therefore has to be
ruled out.  Classical communication between
Alice and Bob, on the other hand, will increase Bob's knowledge of the
\emph{actual} particle number distribution, but it cannot influence the
\emph{average} distribution obtained.

In order to formulate this result precisely, note that any bipartite state
$\phi\in\mc H_N$ subject to SSR can be written as
$\phi=\phi^0\oplus\dots\oplus\phi^N$
with $\phi^n\in\mc H^A_n\otimes\mc H^B_{N-n}$, i.e., as a direct sum of
unnormalized pure states with constant local particle number. Call the
(ordered) \emph{unnormalized} Schmidt coefficients of $\phi^n$ $\bm\lambda^n$.
Then, $\phi$ is characterized up to local (SSR-compatible) unitaries by its
\emph{SSR-ordered Schmidt vector}
$\bm\lambda=(\bm\lambda^0,\dots,\bm\lambda^N)$. 

\begin{theorem}[\cite{footnote:maj-ancillas}]
\label{theorem:majorization}
Let $\phi$, $\psi_i$ be pure states and 
$\bm\lambda$, $\bm\mu_i$ their SSR-ordered Schmidt vectors. Then, 
\begin{equation}
\label{eq:maj-theorem-lhs}
\phi\stackrel{\mathrm{SSR}}{\longrightarrow}\{(p_i,\psi_i)\}
\end{equation}
(i.e., there exists a SSR-compatible conversion strategy)
if and only if 
\begin{equation}
\label{eq:maj-theorem-rhs}
\bm\lambda^n\prec\sum_ip_i\bm\mu_i^n\quad\forall\,n=0,\dots,N.
\end{equation}
\end{theorem}

In order to see the connection to the conversions within the subspaces,
let us re-express~(\ref{eq:maj-theorem-rhs}) by normalizing the
SSR-ordered Schmidt vectors,
$$
\hat{\bm\lambda}^n\prec
\sum_i\underbrace{p_i\frac{\|\bm\mu_i\|}{\|\bm\lambda\|}}_
{=:p_i'}
\hat{\bm\mu}_i^n
\quad\forall\,n=0,\dots,N\,,
$$
where in the following a hat $\hat\cdot$ denotes
the normalized vector. According to the usual majorization result, this
holds iff we can convert
\begin{equation}
\label{eq:maj-theorem-rhs-equiv}
\hat\phi^n\longrightarrow\{p_i',\hat\psi_i^n\}\quad\forall n=0,\dots,N\ .
\end{equation}
Here, $\phi=\phi^0\oplus\dots\oplus\phi^N$ and 
$\psi_i=\psi_i^0\oplus\dots\oplus\psi_i^N$.

\begin{proof}
Exactly as without SSR, the most general strategy consists of
Alice performing a generalized measurement and communicating the result to
Bob, who then applies a unitary operation depending on the measurement
outcome; the proof~\cite{lo:concentrating_ent} can be directly transferred.

We show (\ref{eq:maj-theorem-lhs})$\Leftrightarrow$%
(\ref{eq:maj-theorem-rhs-equiv}).
The proof can be restricted to the case where each
conversion $\phi\rightarrow(p,\psi)$ in~%
(\ref{eq:maj-theorem-lhs}) resp.~(\ref{eq:maj-theorem-rhs-equiv})
can be accomplished by a single POVM operator $M$, i.e., 
$M\phi=\sqrt{p}\psi$---otherwise, 
we can split $\phi\rightarrow(p,\psi)$ into 
$\phi\rightarrow(p_k,\psi)$, $\sum p_k=p$, where each conversion is the
result of \emph{one} of the POVM operators. This can be done as well for the
system of
conversions (\ref{eq:maj-theorem-rhs-equiv}), where we have to split all
subspaces simultaneously (this can be always done by additionally
splitting single POVM
operators into copies of itself).

First, assume that (\ref{eq:maj-theorem-lhs}) holds. Then there exist POVM
operators $M_i=\bigoplus_n M_i^n$ on Alices side 
for which $M_i\phi\cong_B\sqrt{p_i}\psi$ (i.e., up to a unitary on Bob's
side).
Decomposing this into the subspaces in the direct sum, one obtains
$M_i^n\phi^n\cong_B\sqrt{p_i}\psi_i^{n}$ and thus
$$
M_i^n\hat\phi^n\cong_B
\underbrace{
\sqrt{p_i\frac{\bra{\psi_i}\psi_i\rangle}{\bra{\phi}\phi\rangle}}
}_{\equiv \sqrt{p_i'}}
\hat\psi_i^{n}
$$
for all $N$, i.e., the $M_i^n$ accomplish the set of conversions given by
Eq.~(\ref{eq:maj-theorem-rhs-equiv}). Especially, as
$$
\openone=\sum_i\left(\bigoplus_nM_i^n\right)^\dagger
\left(\bigoplus_nM_i^n\right)=
\bigoplus_n\left(\sum_iM_i^{n\dagger}M_i^n\right)\ ,
$$
the $M_i^n$ obey the completeness relation for POVM operators.
As all arguments hold in both directions, this completes the proof.
\end{proof}

\subsection{Variance as a nonlocal monotone}

Let us now formulate an asymptotic version of the previous theorem. It is
known that without SSR for a large number of copies the majorization
criterion converges to the entropic criterion, i.e., the conservation of
the total EoE~\cite{bennett:purestate-ent}. With SSR, the
probability distribution associated to the variation of the local particle
number, $p_n=\sum_ip_i^n$, has to be conserved as well. Asymptotically, this
distribution converges to a
Gaussian which is completely characterized by its mean (which can be
shifted using ancillas) and its variance. Therefore we define

\begin{definition}
For a bipartite pure state $\phi$ shared by A and B,
define the superselection induced variance (SiV)
$$
V(\phi):=4\left[\bra\phi\hat N_A^2\ket\phi-\bra\phi\hat N_A\ket\phi^2\right]\ ,
$$
where $N_A$ is the particle number operator for Alice.  (One could equally
well take $\hat N_B$, as $\hat N_A+\hat N_B=N=\mathrm{const}$.)
\end{definition}
The factor $4$ in the definition normalizes the SiV: $V(\VEPR)=1$.

Let us now show that SiV is really an entanglement
monotone~\cite{vidal:ent_monotone} when SSR are present,
 namely that it cannot be increased on average by SSR-LOCC and
vanishes on separable states.  (On the contrary, note that $V(\phi)=0$
does \emph{not} imply that $\phi$ is separable---this is due to the fact
that there exist two different nonlocal quantities when SSR are present.)
Moreover, SiV is symmetric under interchange of  $A$ and $B$ and additive:
given two subsystems $1$ and $2$ shared by $A$ and $B$,
$V(\phi_1\otimes\phi_2)=V(\phi_1)+V(\phi_2)$, as  can be
readily seen by applying Eq.~(\ref{eq:N_additivity}) to the two
subsystems $1$ and $2$, $\hat N_{A1\,A2}=
\hat N_{A1}\otimes\openone_{A2}+\openone_{A1}\otimes\hat N_{A2}$.

To show the monotonicity of SiV under SSR-LOCC, consider a POVM
measurement $\{M_i^A\}$ on Alice's side. Then, the average SiV after the
application of $\{M_i^A\}$ is given by 
$$
\bar V_M(\phi)=\sum_i
\bra\phi M_i^{A\dagger}\hat N_A^2M_i^A\ket\phi-
   \sum_i\frac{\bra\phi M_i^{A\dagger}\hat N_A M_i^A\ket\phi^2}
	{\bra\phi M_i^{A\dagger}M_i^A\ket\phi}\ .
$$
The first part reduces to $\bra\phi\hat N_A^2\ket\phi$
(using  $[\hat N_A,M_i^A]=0$ and $\sum_iM_i^{A\dagger}M_i^A=\openone$),
while for the second part 
\begin{eqnarray*}
\sum_i\frac{\bra\phi M_i^{A\dagger}\hat N_A M_i^A\ket\phi^2}
	 {\bra\phi M_i^{A\dagger} M_i^A\ket\phi}
	 &\stackrel{\mathrm{(*)}}{\ge}&
    \left(\sum_i\bra\phi M_i^{A\dagger}\hat N_A M_i^A\ket\phi\right)^2\\
    &=&
    \bra\phi\hat N_A\ket\phi^2\ .
\end{eqnarray*}
Here, $\mathrm{(*)}$ has been derived using the Cauchy-Schwarz inequality
\begin{equation}
\left(\sum_iy_i\right)^2=
    \left(\sum_i\sqrt{p_i}\frac{y_i}{\sqrt{p_i}}\right)^2\le
    \sum_i\frac{y_i^2}{p_i}\sum_{i'}p_{i'}\ .
\label{eq:C-S-mod}
\end{equation}
Ancillas leave the result unaffected, as the extra contributions in 
$\bar V_M(\phi)$ originating from $[\hat N,\mc O]=\nu\mc O$ cancel out.

\subsection{Reversible conversion of multiple copies}

The introduction of SiV as a nonlocal monotone was motivated by the
conversion of multiple copies, as it characterizes the joint particle
number distribution. In the following, we will show that
asymptotically SiV and EoE quantify the two complementary
resources which completely characterize biparitite states up to SSR-LOCC.

\begin{theorem}
\label{theorem:EoE-SiV}
In the presence of SSR, there
exists an asymtotically reversible conversion
$$
\ket{\phi}^{\otimes N}\otimes\ket{\hat0}^{\otimes E(\phi)N}\longleftrightarrow
    \sum_n c_n\ket n\ket{N-n}\otimes\CEPR^{\otimes E(\phi)N},
$$
where the coefficients $c_n$ are
distributed Gaussian with SiV
$N\,V(\ket\phi)$. 
\end{theorem}

Note that on the left hand side we have added ancilla states in the
unrestricted ``hat''--basis (cf.~Sec.~\ref{section:definitions}).  The
conversion transfers the entanglement contained in $\ket\phi^{\otimes N}$
to this second register as ``accessible'' entanglement in the form of
$\CEPR$s, while the SiV stays in the first register.

\begin{proof}
First, we restrict ourself to the case of qubits, where
$\ket\phi=\sqrt{p_0}\ket0\ket1+ \sqrt{p_1}\ket1\ket0$. 
We will generalize the result in two steps:
in a first step, we consider \mbox{qu-$d$-its}, where the local basis is
$\{\ket0,\dots,\ket{d-1}\}$, while in a second step we allow for
arbitrary bipartite states, i.e., the local bases might contain several states
with the same particle number.

For the beginning, let us only look at the first register.
Taking $N$ copies of $\ket\phi$, we have
$$
\ket\phi^{\otimes N}=\sum_{\mathbf x}\sqrt{p_0^{n_0}p_1^{n_1}}
\ket{\mathbf x}{\ket{\neg\mathbf x}}\ ,
$$
where the sum is taken over all possible $N$-bit strings $\mathbf x$. 
Here, $n_0$ and $n_1$ are the numbers of zeroes and ones in $\mathbf x$,
respectively, and
$\neg\mathbf x$ denotes the bitwise \textsc{not} of $\mathbf x$.
 This state can be grouped naturally as
\begin{equation}
\ket\phi^{\otimes N}=\sum_{n_0}\sqrt{p_0^{n_0}p_1^{N-n_0}\left(N\atop n_0\right)}
\ket{\chi_{N-n_0,n_0}}\;,
\label{eq:asymp-qubit-by-n}
\end{equation}
where the state $\ket{\chi_{N-n_0,n_0}}\in\mathcal H_{N-n_0}^A\otimes
\mathcal H_{n_0}^B$ is a maximally entangled state with Schmidt number
$\left(N\atop n_0\right)$.

In the following, we show how to transfer the entanglement of
$\ket\phi^{\otimes N}$ to the second register.  Therefore, we have to
break the tensor product structure $\ket\phi^{\otimes N}$ of the first
register and create a new tensor product structure by properly
transferring the entanglement to the second register.  To this end, let us
introduce the concept of typical subspaces~\cite{cover-thomas}. An
$\epsilon$-typical subspace of our Hilbert space is defined as
$\mathcal H_\epsilon=
\bigoplus_{n_0\in\mc S_\epsilon}\mathcal H_{N-n_0}^A\otimes\mathcal H_{n_0}^B$, 
where the $\epsilon$-typical $n_0$ are those lying in 
$\mc S_\epsilon=\{n_0:|n_0/N-p_0|<\epsilon\}$.
It can be shown~\cite{cover-thomas,nielsen-chuang} that projecting
$\ket\phi^{\otimes N}$ onto $\mc H_\epsilon$ gives an error which vanishes
for $N\rightarrow\infty$ such that we can restrict the sum
in~(\ref{eq:asymp-qubit-by-n}) to $n_0\in\mc S_\epsilon$. Then,
\begin{equation}
\label{eq:qubit-typical-lower}
\left(N\atop n_0\right)\ge
\frac{1}{(N+1)^2}2^{NH\left(\frac{n_0}{N}\right)}\ge
2^{N[H(p_0)-K\epsilon]}
\end{equation}
with some $K>0$ holds for all $n_0\in\mc S_\epsilon$ ($\epsilon\ll1$ and
$N\gg1$)~\cite{cover-thomas}; here
$H(p)=H(p,1-p)$ is the Shannon entropy of the probability distribution
$(p,1-p)$.
According to Theorem~\ref{theorem:majorization}, we can transform
$$
\ket{\chi_{N-n_0,n_0}}\rightarrow\frac{1}{\sqrt{E}}
    \sum_{i=1}^E\ket{i_{N-n_0}}_A\ket{i'_{n_0}}_B\ ;\
    E=H(p_0)-K\epsilon
$$
coherently in all subspaces in the restricted sum, where $\ket{i_n}$ are
orthogonal states with $n$ particles.
 Then by local maps 
$\ket{i_n}\ket{\hat0}\mapsto\ket{n}\ket{\hat i}$, where $\ket{\hat i}$ are
orthogonal and $\ket{n}=\ket{1\cdots10\cdots0}$, the entanglement 
$H(p_0)-K\epsilon$ can be transferred to the second register which gives
\begin{equation}
\sum_{n_0\in\mc S_\epsilon}c_{n_0}\ket{N-n_0}\ket{n_0}\otimes
\Big[\ket{01}\ket{10}+\ket{10}\ket{01}\Big]^{\otimes N[H(p_0)-K\epsilon]}\ ,
\label{eq:asym-qubit-factorized}
\end{equation}
where $c_{n_0}=\sqrt{p_0^{n_0}p_1^{N-n_0}\left(N\atop n_0\right)}$.
The sum can be extended to all $n_0$ with high fidelity,
and the $|c_{n_0}|^2$ approach a Gaussian
distribution with variance $Np_0(1-p_0)=V(\phi)/4$. This is the only parameter
characterizing the state~(\ref{eq:asym-qubit-factorized}),
since the mean can be shifted by locally adding
ancillas. As $H(p_0)$ is just $E(\phi)$, this completes the
distillation direction of the proof.

The dilution direction can be proven using the converse
of~(\ref{eq:qubit-typical-lower}),
$$
\left(N\atop n_0\right)\le
2^{NH\left(\frac{n_0}{N}\right)}\le
2^{N[H(p_0)+K\epsilon]}\;,
$$
in an $\epsilon$-typical subspace.
Starting from 
$$
\sum_{n_0\in\mc S_\epsilon}c_{n_0}\ket{N-n_0}\ket{n_0}\otimes
\Big[\ket{01}\ket{10}+\ket{10}\ket{01}\Big]^{\otimes N[H(p_0)+K\epsilon]}\;,
$$
we can transfer the entanglement to the first register and then
(again by Theorem~\ref{theorem:majorization}) reduce the Schmidt
number of each subspace to $\left(N\atop n_0\right)$, obtaining the
projection of $\ket\phi^{\otimes N}$ onto the $\epsilon$-typical subspace, so that 
the dilution works as well. This completes the proof for qubits.

In a first step, we generalize the proof from qubits to $(I+1)$-level systems,
\begin{equation}
\label{eq:multi-state}
\ket\phi=\sum_{i=0}^{I}\sqrt{p_i}\ket{i}\ket{I-i}\;.
\end{equation}
(Note that the coefficients can be made positive by local operations.)
Again, for $N$ copies of $\ket\phi$, an $\epsilon$-typical subspace can be
defined by restricting the number $n_i$ of occurences of the state
$\ket{i}\ket{I-i}$ in the product by $|n_i/N-p_i|<\epsilon$ for all $i$.
Projecting the state onto an $\epsilon$-typical subspace again only yields 
a vanishingly small error, and the Schmidt number of the states with fixed
numbers $(n_0,\dots,n_I)$ is given by the multinomial coefficient
$\left(N\atop n_0\ \cdots\ n_I\right)$ and obeys the
bounds~\cite{cover-thomas}
$$
2^{N[E(\phi)-K\epsilon]}\le
\left(N\atop n_0\ \cdots\ n_I\right)\le
2^{N[E(\phi)+K\epsilon]}\;.
$$
Thus, it is possible to extract the entanglement $E(\phi)$
reversibly. Yet, there are several possible configurations
$(n_0,\dots,n_I)$ which yield the same local particle number
$n=\sum_iin_i$ such that there is still some entanglement left in each
subspace. But as for $N$ copies of $\ket\phi$ the number of these
configurations is bounded by $N^I$, this entanglement is 
logarithmic in $N$ and can be removed reversibly. Therefore,
we can reversibly transform $\ket\phi^{\otimes
N}\otimes\ket{\hat0}^{\otimes NE(\phi)}$ into
\begin{equation}
\sum c_{n}\ket{n}\ket{IN-n}\otimes
\Big[\ket{01}\ket{10}+\ket{10}\ket{01}\Big]^{\otimes NE(\phi)}\;,
\end{equation}
where the $c_{n}$ are given by the sum over all 
coefficients for which the particle number on Alice's side is $n$,
$$
c_n=\sqrt{
    \sum_{\sum_i in_i=n\atop\sum_i n_i=N}
    p_0^{n_0}\cdots p_I^{n_I}\left(N\atop n_0\ \cdots\ n_I\right)
}\quad.
$$

%%%%%%%%%%%%%%%%%%%%%%%%%%%%%%%%%%%
%%%%% FIG 1 %[purestate-resources]%
%%%%%%%%%%%%%%%%%%%%%%%%%%%%%%%%%%%
\begin{figure}[t]
\includegraphics[width=0.9\columnwidth]{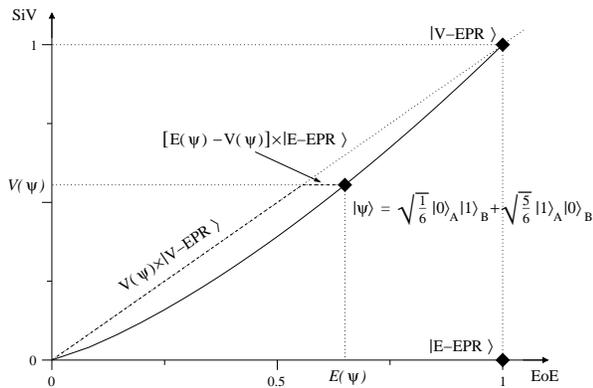}
\caption{
\label{fig:purestate-resources}
Characterization of pure qubit states in an $E$-$V$ diagram. All the
states reside on the solid curve; asymptotically, any state can be
converted into $V(\psi)$ copies of a $\VEPR$ and $E(\psi)-V(\psi)$ of a
$\CEPR$.
}
\end{figure}

It remains to be shown that the $|c_n|^2$ approach a Gaussian distribution.
As long as all $p_i\ne0$, this can be shown by expanding each $n_i$ within the 
typical subspace as $n_i=N(p_i+\delta_i)$ with $\delta_i<\epsilon$. This will
work fine whenever $Np_i\gg1$ and $\epsilon\ll p_i$ for all $i$. Yet, this 
condition cannot be satisfied if $p_i=0$ for some $i$. This might (but
need not!) lead to a periodic gap in the distribution of the $|c_n|^2$,
e.g., for $I=2$, $p_0=p_2=1/2$. In that case, $|c_n|^2=0$ for all odd $n$.  

In principle, such a gap has to be considered as a third nonlocal
characteristic of a bipartite state. Still, it can be removed easily. In
the example given above the gap is readily removed by adding \emph{one} $\VEPR$,
such that the fraction of $\VEPR$ per copies of $\ket\phi$ vanishes. By further
adding an $\CEPR$ (those are obtained anyway in the distillation) the 
$\VEPR$ can be re-obtained---it therefore merely acts as a catalyst, 
``freeing'' the subspaces with odd particle number. 

The generalization to an arbitrary state is straightforward. Take
\begin{equation}
\label{eq:asym_mostgeneral_outer}
\ket\phi=\sum_{i=0}^{I}\sqrt{p_i}\ket{\psi_{i,I-i}}\;,
\end{equation}
where $\ket{\psi_{i,I-i}}\in\mathcal H_i^A\otimes\mathcal H_{I-i}^B$ might
themselves be entangled states. 
Applying the concept of typical subspaces to
Eq.~(\ref{eq:asym_mostgeneral_outer}), we find that the number of
occurences of each $\ket{\psi_{i,I-i}}$ in the typical subspace is bounded
by $(p_i\pm\epsilon)N$, and thus the entanglement
$E(\psi_{i,I-i})$ contained in these states---which is already
``accessible entanglement''---can be extracted reversibly.
The remaining state is of the type
of Eq.~(\ref{eq:multi-state}) (with
the same coefficients $p_i$) and thus can be transformed reversibly into a
Gaussian distributed state with width $NV(\phi)$ and $NH(p_0,\dots,p_I)$
ebits of entanglement. It can be checked easily that the total number of
Bell pairs is $NE(\phi)$.
\end{proof}

%%%%%%%%%%%%%%%%%%%%%%%%%%%%%%%%%%%
%%%%% FIG 2 %[qutrit-diagram]%%%%%%
%%%%%%%%%%%%%%%%%%%%%%%%%%%%%%%%%%%
\begin{figure}[t]
\includegraphics[width=0.9\columnwidth]{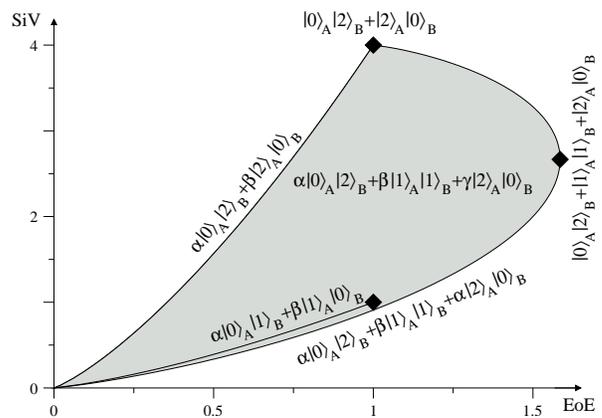}
\caption{
\label{fig:qutrit-diagram}
$E$-$V$ diagram for qutrits, where the boundary states and the extremal
states are given. The possible states reside in the gray area, the solid
line within this area is the subset realizable by qubits.
}
\end{figure}

For qubits, Theorem~\ref{theorem:EoE-SiV} can be re-expressed.
\begin{corollary}
\label{theorem:qubit-resources}
For bipartite qubit states $\ket\phi$, 
$$
\ket\phi\longleftrightarrow\CEPR^{\otimes[E(\phi)-V(\phi)]}\VEPR^{\otimes V(\phi)}
$$
in the asymptotic limit.
\end{corollary}
This can be shown by applying Theorem~\ref{theorem:EoE-SiV} twice,
together with $E(\phi)\ge V(\phi)$ (which only holds for qubits).

Fig.~\ref{fig:purestate-resources} illustrates this characterization 
of states in the $E$-$V$ diagram.
Fig.~\ref{fig:qutrit-diagram} shows the $E$-$V$ diagram for qutrits, which
is considerable more complex. The bounds are given by the states with
highest variance $\alpha\ket0_A\ket2_B+\beta\ket2_A\ket0_B$ and the
states with highest entanglement
$\alpha\ket0_A\ket2_B+\beta\ket1_A\ket1_A+\alpha\ket0_A\ket2_B$.  A
decomposition as in the Corollary is still
possible if one replaces the $\VEPR$ by 
$\ket0_A\ket2_B+\ket2_A\ket0_B$ which has maximal variance.

%%%%%%%%%%%%%%%%%%%%%%%%%%%%%%%%%%%%%%%%%%%%%%%%%%%%%%%%%%%%%%%%%%%%%
%%%%%%%%%%%%%%%%%%%%%%%%%%%%%%%%%%%%%%%%%%%%%%%%%%%%%%%%%%%%%%%%%%%%%
\section{Mixed states in the presence of superselection rules
\label{section:mixed}}
%%%%%%%%%%%%%%%%%%%%%%%%%%%%%%%%%%%%%%%%%%%%%%%%%%%%%%%%%%%%%%%%%%%%%
%%%%%%%%%%%%%%%%%%%%%%%%%%%%%%%%%%%%%%%%%%%%%%%%%%%%%%%%%%%%%%%%%%%%%

\subsection{Introduction}

In the following section, we consider mixed states. We will show how the
concepts of EoE and SiV as two complementary resources can be extended to
mixed states, and discuss the connection with normal (SSR-free)
entanglement measures.

Let us start by introducing a particularly interesting mixed state,
\begin{equation}
\rho_\mathrm{sep}=\frac14\left(
	    \begin{array}{cccc}1\\&1&1\\&1&1\\&&&1\end{array} 
	\right)
\label{eq:rhoSep}
\end{equation}
in the basis
$\{\ket0_A\ket0_B,\ket0_A\ket1_B,\ket1_A\ket0_B,\ket1_A\ket1_B\}$.
This state has first been considered in~\cite{verstraete:SSR-datahiding},
where it was shown that it is separable but nonlocal. Namely, it
can be obtained by mixing 
$(\ket0_A+\omega\ket1_A)(\ket0_B+\omega\ket1_B)$ for
$\omega\in\{1,i,-1,-i\}$ with equal probabilities and therefore does not
contain EoE. On the other hand, it
is easy to see that there is no decomposition of $\rhoSep$ which is
separable \emph{and} compatible with the superselection rule, i.e., it
cannot be created locally. Clearly, this can not happen with pure states.

Considering the results of the preceding section, it is natural to assume
that $\rhoSep$ contains SiV but no EoE. In order to give quantitative
meaning to such statements, we discuss two genuine extensions of
nonlocal quantities to mixed states, defined by the asymptotic amount of
pure state resources which are needed to create them and
which can be extracted again.

\subsection{Formation of mixed states \label{section:mixed-formation}}

Let us start with the creation of mixed state in the presence of SSR.
Similar to the normal case~\cite{bennett:form-rec-hashing}, we define:

\begin{definition}
The entanglement of formation and the variance of formation in the
presence of superselection rules are defined as
$$
E_F^\SSR(\rho)=\min_{\{p_i,\psi_i\}}\sum_ip_iE(\psi_i)
$$
and
$$
V_F^\SSR(\rho)=\min_{\{p_i,\psi_i\}}\sum_ip_iV(\psi_i)\ ,
$$
respectively. The minimum is taken over all possible decompositions of
$\rho$, where the $\psi_i$ have to obey the SSR (i.e., they all have
constant particle number).

The entanglement cost~\cite{hayden:ent_cost} and the variance cost in the
presence of superselection rules are accordingly defined as the
regularized versions of $E_F^\SSR$ and $V_F^\SSR$,
$$
E_c^\SSR(\rho)=\lim_{N\rightarrow\infty}\frac{E_F^\SSR(\rho^{\otimes N})}{N}
$$
and
$$
V_c^\SSR(\rho)=\lim_{N\rightarrow\infty}\frac{E_F^\SSR(\rho^{\otimes N})}{N}\ .
$$
\end{definition}
These definitions make sense, as they quantify the nonlocal resources we
need at least to prepare the state $\rho$ with SSR~\cite{hayden:ent_cost}.

As shown at the beginning of the section there exist states which do not
contain any entanglement yet are nonlocal, as $\rhoSep$
[Eq.~(\ref{eq:rhoSep})]. One easily finds that 
$E_F^\SSR(\rho_\mathrm{sep})=1/2$, $V_F^\SSR(\rho_\mathrm{sep})=1/2$, as each of
the subblocks in $\rho_\mathrm{sep}$ has to be created separately. 
On the other hand, it seems reasonable to assume that $\rhoSep$ can be
prepared asymptotically without using entanglement. In the following, we
prove an even stronger result: asymptotically, the entanglement needed to
create any state $\rho$ is just the entanglement needed without SSR.
\begin{theorem}
For any $\rho$ with bounded maximal particle number,
$$
E_c^\SSR(\rho)=E_c(\rho)\ ,
$$
i.e., the entanglement cost with SSR is the entanglement cost without SSR.
\end{theorem}

\begin{proof}
Consider a mixed state $\sigma$ compatible with the SSR 
and let $\sum_ip_i\ket{\psi_i}\bra{\psi_i}=\sigma$ be the optimal
decomposition without SSR, i.e., $E_F(\rho)=\sum_ip_iE(\psi_i)$. 
Clearly, this decomposition need not 
obey the SSR, but we can use it to constuct a compatible decomposition
with vanishing overhead.
From~(\ref{eq:OisPnOPn}),
$\sigma=\sum_{n=0}^NP_n\sigma P_n$, where $P_n$ is the projector onto the
subspace with \emph{totally} $n$ particles and $N$ the maximum total
particle number in $\sigma$; therefore,
$$
\sigma=\sum_{n,i}p_ip_{i,n}\frac{P_n\ket{\psi_i}\bra{\psi_i}P_n}{p_{i,n}}
$$
with $p_{i,n}=\bra{\psi_i}P_n\ket{\psi_i}$ is a decomposition of $\sigma$
which is compatible with the SSR. For any $\ket\psi$ with at most $N$
particles, it holds that the measurement of the total particle number
creates at most $\log(N+1)$ entanglement on average,
\begin{equation}
\label{eq:P_n-log-creation}
\sum_n\bra\psi P_n\ket\psi 
    E\left(\frac{P_n\ket\psi}{\sqrt{\bra\psi P_n\ket\psi}}\right)
    \le
    E(\ket\psi)+\log(N+1)
\end{equation}
(the proof is given in the appendix),
and with $\sigma=\rho^{\otimes M}$, the claim follows.
\end{proof}

Note that this also implies that $E_F^\SSR$ is not
additive~\cite{wiseman:SSR:M};
$E_F^\SSR(\rhoSep^{\otimes N})$, e.g., grows at most logarithmically.

Let us now consider $V_F^\SSR$ and $V_c^\SSR$. As expected, the
entanglement cost of $\rhoSep$ vanishes. But as $\rhoSep$ still contains
some kind of nonlocality, it is natural to assume that its variance cost
is strictly nonzero. In the following, we prove a more general
result, namely that $V_F^\SSR$ is additive on all states which are a
direct sum of pure states (i.e., $\rho$ is block-diagonal and each block
is a pure state); this holds, e.g., for $\rhoSep$.

\begin{theorem}
\label{theorem:VF-additive}
Let $\rho=\bigoplus_ip_i\ket{\phi_i}\bra{\phi_i}$,
$\sigma=\bigoplus_jq_j\ket{\psi_j}\bra{\psi_j}$, where
$\sum_ip_i=\sum_jq_j=1$. Then
$$
V_F^\SSR(\rho\otimes\sigma)=V_F^\SSR(\rho)+V_F^\SSR(\sigma)\ .
$$
\end{theorem}
\begin{proof}
\begin{eqnarray*}
V_F^\SSR(\rho\otimes\sigma)&=& 
    V_F^\SSR\left(
        \bigoplus_{i,j}p_iq_j\ket{\phi_i}\bra{\phi_i}\otimes\ket{\psi_j}\bra{\psi_j}
    \right) 
\\
&\stackrel{(*)}=&\sum_{i,j}p_iq_jV(\phi_i\otimes\psi_j)
\\
&=&\sum_ip_iV(\phi_i)+\sum_jq_jV(\psi_j)
\\
&\stackrel{(*)}=&V_F^\SSR(\rho)+V_F^\SSR(\sigma)\ ,
\end{eqnarray*}
where in $(*)$ we used the equality
$V_F^\SSR(\bigoplus_i r_i\ket{\chi_i}\bra{\chi_i})=\sum_ir_iV(\chi_i)$
with $\sum_ir_i=1$. As subadditivity is clear from the convexity of
$V_F^\SSR$, we only have to show superadditivity. For an arbitrary
decomposition $\ket{\zeta_j}=\sum_iu_{ji}\sqrt{r_i}\ket{\chi_i}$ of 
$\bigoplus_ir_i\ket{\chi_i}\bra{\chi_i}=\sum_j\ket{\zeta_j}\bra{\zeta_j}$ 
[with an isometry $(u_{ji})$], this follows from 
\begin{eqnarray*}
\sum_j\frac{\bra{\zeta_j}\hat N_A\ket{\zeta_j}^2}{\bra{\zeta_j}\zeta_j\rangle}
    \stackrel{(\mathrm{a})}{=}
\sum_j\frac{(\sum_iu_{ji}^*u_{ji}p_i\bra{\chi_i}\hat N_A\ket{\chi_i})^2}
	{\sum_iu_{ji}^*u_{ji}p_i}\qquad\\
    \qquad\stackrel{(\mathrm b)}{\le}
\sum_{i,j}
    \frac{(u_{ji}^*u_{ji}p_i\bra{\chi_i}\hat N_A\ket{\chi_i})^2}
	{u_{ji}^*u_{ji}p_i}
    \stackrel{(\mathrm c)}{=}
\sum_{i}p_i
    \bra{\chi_i}\hat N_A\ket{\chi_i}^2\ .
\end{eqnarray*}
Here, we used $(\mathrm a)$ 
$\bra{\chi_i}\chi_{i'}\rangle=\delta_{ii'}$,
$\bra{\chi_i}\hat N_A\ket{\chi_{i'}}\propto\delta_{ii'}$;
$(\mathrm b)$~Eq.~(\ref{eq:C-S-mod});
$(\mathrm c)$ $\sum_j u_{ji}^*u_{ji}=1$.
\end{proof}

While it seems plausible that $V_F^\SSR$ is additive on all states and
we did not find any counterexamples, this is apparently hard to prove.
Let us note that unlike for
$E_F$, the additivity of $V_F^\SSR$ is probably not related to its
superadditivity. A counterexample for the superadditivity of $V_F^\SSR$
can easily be found~\cite{footnote1},
and the
direct equivalence proof of Pomeransky~\cite{pomeransky:EoF-add-superadd}
cannot be transferred to SiV
due to the different structure of the nonlinearity.

\subsection{Formation of qubits
    \label{section:qubit-EF-VF}}

In the following, we compute explicit formulas for 
$E_F^\SSR$ and $V_F^\SSR$ of qubits. 
A general bipartite two-qubit state subject to SSR is given by
$$
\rho=\left(\begin{array}{cccc}
    w_{00}\\
    &w_{01}&\gamma\\
    &\gamma&w_{10}\\
    &&&w_{11}
\end{array}\right)\ ,
$$
where $\gamma\ge0$ (this can be achieved by local unitaries).
Using the results of Wootters~\cite{wootters:qubit-EoF},
we find $E_F(\rho)=\mc E(C)$,
where $\mc E(C)=H(1/2+\sqrt{1-C^2}/2)$, $H$ is the binary entropy,
and the concurrence $C\equiv C(\rho)$ can be computed as
$$
C=\max(0,2\gamma-2\sqrt{w_{00}w_{11}})\ .
$$
With SSR, $\rho$ has to be built subspace by subspace, where the
one-particle subspace $\rho_1$ is the only one which might be entangled.
The concurrence for $\rho_1/\Tr[\rho_1]$ is
$$
\bar C=2\gamma/p
$$
with $p=w_{01}+w_{10}=\Tr[\rho_1]$, and thus
$$
E_F^\SSR(\rho)=p\mc E(\bar C)\ .
$$
The relation between the normal concurrence $C$  and the SSR-concurrence
$\bar C$ is given by the bounds $p\bar C-(1-p)\le C\le p\bar C$, i.e.,
$E_F$ and $E_F^\SSR$ are not completely independent. As $\mc E$ is
concave, $E_F\le E_F^\SSR$, as necessary.

An optimal decomposition of $\rho_1$ can be found as follows. Define $s$
as a root of $\bar C/2=\sqrt{s(1-s)}$. Then, $\rho_1$ can be
written as a mixture of $\sqrt{s}\ket{01}+\sqrt{1-s}\ket{10}$ and
$\sqrt{1-s}\ket{01}+\sqrt{s}\ket{10}$, and both have the desired EoE. 

The same decomposition gives the optimal variance as well 
(note that this only holds for qubits).
Therefore, observe that both states have SiV $4s(1-s)=\bar C^2$, i.e., 
$\bar C^2$ is an upper bound for $V_F^\SSR(\rho)$, and for pure states
equality holds. On the other hand, $\bar C^2$ is convex: for any
one-particle subblock $\rho_1=p\sigma+(1-p)\sigma'$ with
off-diagonal elements $v=pw+(1-p)w'$ it holds that $v^2\le pw^2+(1-p)w'^2$.
Therefore equality holds, and
\begin{equation}
\label{eq:qubit-VoF}
V_F^\SSR(\rho)=p\bar C^2\ .
\end{equation}

%%%%%%%%%%%%%%%%%%%%%%%%%%%%%%%%%%%
%%%%% FIG 3 %[qubit-form]%%%%%%%%%%
%%%%%%%%%%%%%%%%%%%%%%%%%%%%%%%%%%%
\begin{figure}[t]
\includegraphics[width=0.9\columnwidth]{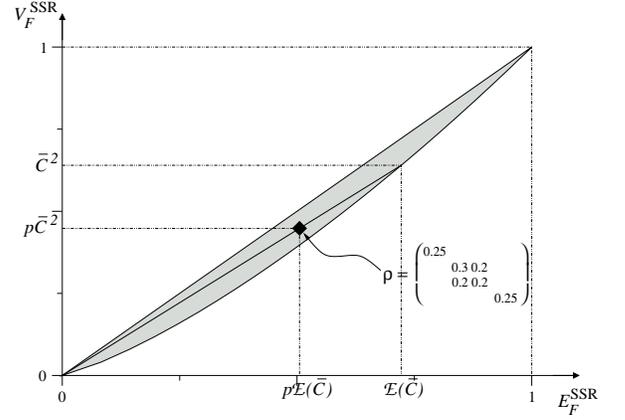}
\caption{
\label{fig:qubit-form}
Relation of $p$, $\bar C$, $E_F^\SSR$ and $V_F^\SSR$ (see
Section~\ref{section:qubit-EF-VF}).
The gray area gives the allowed range of $E_F^\SSR$ and $V_F^\SSR$ for
qubits. The lower bound is obtained by plotting $\mc E(\bar C)$ vs.\ $\bar
C^2$. The point characterizing a mixed state $\rho$ can be found by
dividing the line between the origin and the point $(\mc E(\bar C),\bar C^2)$ 
located on the boundary at the ratio of $p:1-p$.
}
\end{figure}

Thus, with respect to formation $1\times1$ qubit states 
are characterized by two parameters: the weight of $\rho_1$,
$p$, and the concurrence of $\rho_1$, $\bar C$. It can be checked easily
that $0\le p,\bar C\le 1$ in order for $\rho$ to be positive. 
A necessary condition for separable states is $\bar C\le (1-p)/p$ (this
is tight, but $p$, $\bar C$ do not tell everything about separability,
cf.\ the inequality relating $C$, $\bar C$ given above).

\begin{figure}[t]
\includegraphics[width=0.9\columnwidth]{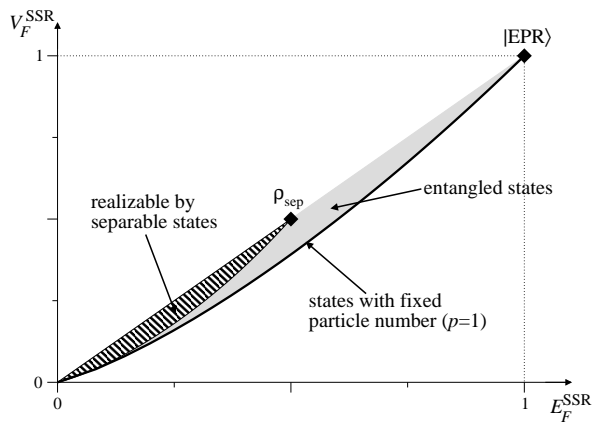}
\caption{
\label{fig:qubit-form-regions}
Different regions of mixed states in the $E$-$V$ diagram. The solid line
corresponds to the states with fixed total particle number, i.e., $p=1$.
Separable states have to stay within the dashed area (although there exist
entangled states as well). Note that $\rhoSep$ [Eq.~\ref{eq:rhoSep}] is
the ``extremal separable state''.
}
\end{figure}

Fig.~\ref{fig:qubit-form} shows how  for a particular
state $p$ and $\bar C$ can be determined from the $E$-$V$ diagram, and
Fig.~\ref{fig:qubit-form-regions} gives a ``phase diagram'' for mixed
states.

\subsection{Distillation of nonlocal resources}

In the following, we consider the problem complementary to formation:
given a mixed state, is it possible to \emph{distill} the nonlocal
resources contained in this state? This distilled state could then be used
to perform some nonlocal task as teleportation with high fidelity.
Naturally, there exist two types of distillation protocols:
the first one aims to increase the fidelity of the states being distilled
with the final state, while
the second returns the target state \emph{itself} with some finite yield in
the asymptotic limit.

In the following, we will focus on qubits. Without SSR, it has been shown
for both types of distillation how they can be implemented: in the
so-called recurrence protocol~\cite{bennett:recurrence}, both parties
apply an \textsc{xor}  operation (a bilateral \textsc{xor} or
\textsc{bxor}) to two copies of the state and then measure one of them;
thus, on average they increase their knwoledge about the second.  Hashing
protocols~\cite{bennett:form-rec-hashing}, on the contrary, are aimed to
asymptotically return a finite yield of pure states: by subsequent
application of \textsc{bxor} operations partial information about the
states can be collected in a subset which is then measured; by the law of
large numbers, this partial information asymptotically fully determines
the remaining states.

%%%%%%%%%%%%%%%%%%%%%%%%%%%%%%%%%%%
%%%%% FIG 4 %[qubit-form-regions]%%
%%%%%%%%%%%%%%%%%%%%%%%%%%%%%%%%%%%

The presence of superselection rules inposes severe restrictions on
distillation procedures. It has been shown, e.g., that the entanglement
contained in \emph{one} copy of a $\VEPR$ cannot be accessed, while for
multiple copies of $\VEPR$, all entanglement up to a logarithmic amount
can be used~\cite{wiseman:SSR:M}, as also follows from
Thm.~\ref{theorem:EoE-SiV}. The central problem in distilling states
containing SiV is that the \textsc{bxor} operation is ruled out by the
SSR, and there is no adequate replacement. One way to overcome this
problem is to use a third copy of the state as a (imperfect) shared
reference frame and to construct a three-copy protocol which
probabilistically implements \textsc{bxor}. Indeed, we will show that one
needs three-copy protocols to distill both EoE and SiV. Unfortunately,
this is of no use for the implementation of hashing protocols, as the
errors of the \textsc{bxor}-approximation accumulate, and each
\textsc{bxor} uses up the reference frame copy of the state whereas hashing
would require $O(N^2)$ \textsc{bxor} operations.

The existence of two distinct resources makes the field of distillation
much more rich: there will occur trade-offs between the two resources in
distillation, and one might even think of \emph{spending} one resource to
distill the other. For instance, we will show that it is possible to
distill all separable but nonlocal states towards
$\rhoSep$~[Eq.~(\ref{eq:rhoSep})], and in turn, if one adds some
entanglement, all the SiV contained in $\rhoSep$ can be converted to a
$\VEPR$.

\subsubsection{Reduction to standard states}

To simplify analysis, in~\cite{bennett:form-rec-hashing} the distillation
of qubits has been considered for a standard form, namely Bell-diagonal
states; any state can be made Bell-diagonal by LOCC.  Yet, these
operations are ruled out by the SSR, so that we have to introduce a
different normal form. Therefore, consider a general bipartite qubit
state with SSR
\begin{equation}
\label{eq:gen-mixed-qubitstate}
\rho=\left(\begin{array}{cccc}
    w_{00}\\
    &w_{01}&\gamma\\
    &\gamma^*&w_{10}\\
    &&&w_{11}
\end{array}\right)\ ,
\end{equation}
where $w_{ij}\ge0$ and $\gamma\ge 0$ (the latter can be accomplished by
local unitaries). By local filtering operations~\cite{verstraete:localfiltering}
$F_A\propto \sqrt[4]{w_{10}w_{11}}\ket0\bra0+\sqrt[4]{w_{00}w_{01}}\ket1\bra1$,
$F_B\propto \sqrt[4]{w_{01}w_{11}}\ket0\bra0+\sqrt[4]{w_{00}w_{10}}\ket1\bra1$,
this can be transformed probabilistically to
\begin{equation}
\label{eq:standard-mixed-qubitstate}
\tilde\rho=\frac{1}{2(1+w)}
\left(\begin{array}{cccc}
    w\\&1&v\\&v&1\\&&&w
\end{array}\right)\ .
\end{equation}
Here, $w=\sqrt{\frac{w_{00}w_{11}}{w_{01}w_{10}}}$ and
$v=\frac{|\gamma|}{\sqrt{w_{01}w_{10}}}$. In the following, we will call
this the \emph{standard form} $\tilde\rho$ of a two-qubit state $\rho$ and
only consider states of this type. The standard form is Bell-diagonal and 
unique for each $\rho$, and by the reverse POVM
$F_A'\propto F_A^{-1}$, $F_B'\propto F_B^{-1}$, $\tilde\rho$ is converted
back to $\rho$. Thus, any state can be transformed probabilistically to
its standard form and back by LOCC, and therefore the standard form of
states containing EoE and SiV still contains EoE and
SiV~\cite{footnote:stdform-special}.

Note that the two parameters $(w,v)$ describing the standard form 
$\tilde\rho$ are directly related to $(p,\bar C)$ used to
characterize $E_F^\SSR(\tilde\rho)$ and $V_F^\SSR(\tilde\rho)$
in Section~\ref{section:qubit-EF-VF}: $v=\bar C$ and $w=1/p-1$.

\subsubsection{Distilling entanglement}

Let us first demonstrate that it is possible to distill all entangled
qubit states, as it is the case without SSR. Therefore, take two copies of 
an arbitrary state $\rho$ in its standard form
$\tilde\rho$~[Eq.~(\ref{eq:standard-mixed-qubitstate})]
and project locally onto the one-particle subspaces. The
resulting state in the $\{\ket{\hat0},\ket{\hat1}\}$-basis is
$$
\hat\rho=\left(\begin{array}{cccc}w^2\\&1&v^2\\&v^2&1\\&&&w^2\end{array}\right)\ .
$$
Obviously, $\hat\rho$ is entangled iff $\tilde\rho$ is entangled iff
$\rho$ is entangled, and as $\hat\rho$ has constant local particle number, it can
be distilled as usual~\cite{bennett:recurrence,bennett:form-rec-hashing}.
Therefore, it is possible to distill any entangled two-qubit state if we
do not care about its SiV. Even more, if we have an infinite amount of SiV
available, we can distill with the same rate as without SSR by using the
SiV as a perfect reference frame.

\subsubsection{Distilling separable states}

Clearly, the SiV contained in separable but nonlocal states cannot be
distilled, as pure states with SiV always contain entanglement. One
solution to this problem is to distill towards
$\rhoSep$~[Eq.~(\ref{eq:rhoSep})]; 
we will show how
this can be done (and why $\rhoSep$ is a good choice) in the next
subsection.  Alternatively, one might try to add entanglement (e.g.,
$\CEPR$s) and then distill the SiV of separable states to $\VEPR$s.

In the following, we show how 
$\rhoSep\otimes\CEPR\bra{\mbox{\rm{E-EPR}}}$ can be transformed to a $\VEPR$ with
probability $1/2$, thereby distilling all the SiV contained in $\rhoSep$
to a pure state.
Clearly,  $\VEPR$ can be obtained from an
$\CEPR=\ket{01}\ket{10}+\ket{10}\ket{01}$ by applying a \textsc{bxor}
operation, but this is ruled out by the SSR. The idea in the following is
to use the SiV contained in $\rhoSep$ as a shared reference frame in order
to carry out the \textsc{bxor} operation probabilistically. 
In order to see how this works, write $\rhoSep$ as a mixture of 
$(\ket0+\omega\ket1)_A(\ket0+\omega\ket1)_B$ over all $\omega=e^{i\phi}$. If
we manage to project the total state onto subspaces where $\omega$ 
simply gives  a global phase, we can make use of the SiV of $\rhoSep$.
Therefore, start with the state 
$\CEPR\bra{\mbox{E-EPR}}\otimes\rho_\mathrm{sep}$
which can be written as a mixture of the states
\begin{eqnarray*}
\ket{\psi_0}&\propto&\ket{010}\ket{100}+\ket{100}\ket{010}\ ,\\
\ket{\psi_1}&\propto&
    \ket{010}\ket{101}+\ket{100}\ket{011}+\ket{011}\ket{100}+\ket{101}\ket{010}\ ,\\
\ket{\psi_2}&\propto&\ket{011}\ket{101}+\ket{101}\ket{011}
\end{eqnarray*}
with probabilities $1/4$, $1/2$, and $1/4$. 
Clearly, there is no measurement which tells us which $\ket{\psi_i}$ 
we actually have without either destroying the entanglement
contained in $\ket{\psi_0}$ and/or $\ket{\psi_2}$ or the
variance contained in $\ket{\psi_1}$. 
As we want to extract the variance, we have to sacrifice the EoE of
$\ket{\psi_{0,2}}$: both parties do a projective measurement
onto the subspaces spanned by $\{\ket{010},\ket{101}\}$ and
$\{\ket{100},\ket{011}\}$.
If the measurement outcomes match, Alice and Bob share a known state with
EoE and SiV $1$ which can be converted to a $\VEPR$; otherwise, 
the entanglement is lost.  Both cases are equally likely, and thus
the average yield
of SiV is $1/2=V_F^\SSR(\rho_\mathrm{sep})$ which is optimal.
On the other hand, we had to sacrifice half of the entanglement---there is
a trade-off between the two resources. 

The procedure described above can be generalized to arbitrary states,
where it allows to distill the one-particle subblock. Note that if $\rho$
is entangled, the required $\CEPR$s can be distilled from $\rho$ itself.

\subsubsection{Recurrence protocols}

In the following, we will look for protocols which allow to distill EoE
\emph{and} SiV. Particularly, we would like to have a protocol which
allows to concentrate the SiV contained in separable states. As already
mentioned at the beginning of the section, the usual recurrence protocols
cannot be applied as \textsc{bxor} cannot be implemented.
(In fact, it is not even possible to find an operation doing
a comparable job, i.e., computing the parity, only for
$\ket0\ket1\pm\ket1\ket0$.) Yet, similar to the preceding subsection we
can use a third copy as a shared reference frame which allows to implement
the desired recurrence operation in a probabilistic way. Indeed, we will
see that three-copy protocols suffice for all distillation tasks.

General $N$-copy recurrence protocols can be represented by local POVM operators
which act on $N$ qubits and leave one qubit (i.e., $2\times2^N$
matrices). These operators must be realizable by SSR-compatible
operations, i.e., by an $N$-qubit POVM, followed by a measurement of all
but one qubit (omitting the measurement decreases our information about
the state and thus does not help). Therefore, the POVM operators must have
the shape of two adjacent rows of SSR-compatible 
$N$-qubit operations~[Eq.~(\ref{eq:OisPnOPn})];
except normalization, this is the only condition.

%%%%%%%%%%%%%%%%%%%%%%%%%%%%%%%%%%%
%%%%% FIG 5 %[distill0-3]%%%%%%%%%%
%%%%%%%%%%%%%%%%%%%%%%%%%%%%%%%%%%%
\begin{figure}
\includegraphics[width=0.9\columnwidth]{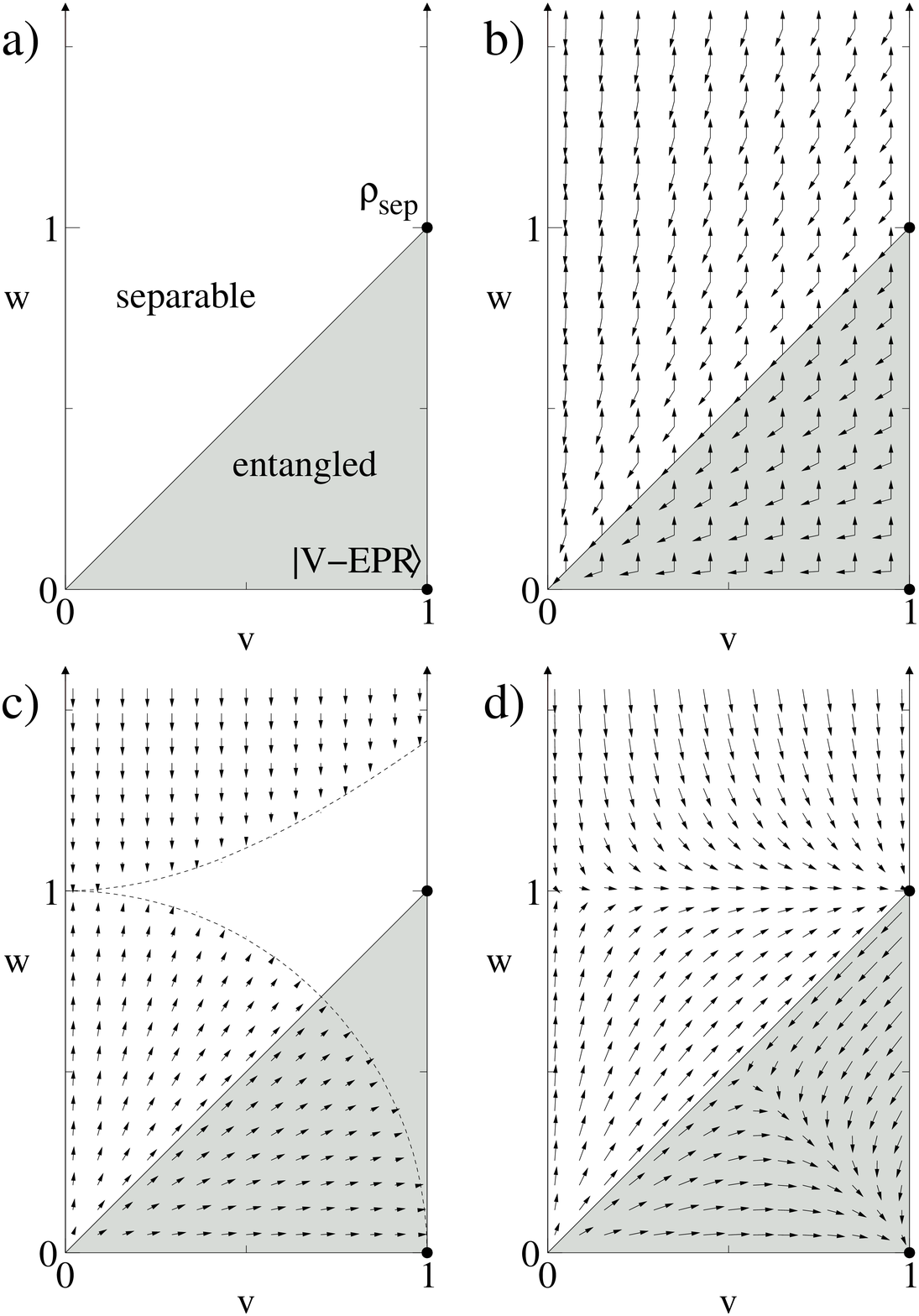}
\caption{
\label{fig:distill0-3}
\textbf{a)}
Diagram characterizing mixed states according to their standard form.
The entangled states are exactly those in the gray area.
\ \textbf{b)}
Transformations possible by one-copy operations: $w$ can be increased, or
$w$ and $v$ can be decreased simultaneously. Thereby, the $\VEPR$ can be
transformed to any state, while $\rhoSep$ can generate any separable
state.
\ \textbf{c)}
Additional transformations realizable by two-copy recurrence protocols.
Thereby, it is not possible to reach $\rhoSep$ or $\VEPR$.
\ \textbf{d)}
Three-copy protocols allow to distill all separable states towards
$\rhoSep$ and all entangled ones towards $\VEPR$.
}
\end{figure}

Possible protocols are illustrated in Fig.~\ref{fig:distill0-3}. Any state
can be brought to  standard form Eq.~(\ref{eq:standard-mixed-qubitstate}) by
local filtering operations and can be parametrized by a tuple $(v,w)$,
$0\le v\le1$, $0\le w$; the states with $v>w$ are entangled
(Fig.~\ref{fig:distill0-3}a). 

Given a \emph{single} copy of $\tilde\rho(v,w)$, Alice and Bob can either
increase $w$ (by adding $\ket{00}\bra{00}+\ket{11}\bra{11}$)  or decrease
$v$ and $w$ by the same fraction (by adding
$\ket{01}\bra{01}+\ket{10}\bra{10}$), and anything inbetween, as is
illustrated in Fig.~\ref{fig:distill0-3}b. Obviously, $\rhoSep$ can be
transformed to any other separable state deterministically---therefore, it
is indeed \emph{the} standard separable but nonlocal state, as an EPR is
for entanglement.

Let us turn our attention to two-copy protocols. As the output state will
not necessarily have standard form, we have to include filtering in the
local POVM operators which restricts their degrees of freedom to
one complex number each, so that it is easy to check that the 
best protocols are given by 
$$
M_A=M_B\propto\left(\begin{array}{cccc}1&0&0&0\\0&1&1&0\end{array}\right)
$$
and
$$
M_A'\propto\left(\begin{array}{cccc}1&0&0&0\\0&1&1&0\end{array}\right)
\ ,\quad
M_B'\propto\left(\begin{array}{cccc}0&1&1&0\\0&0&0&1\end{array}\right)\ .
$$
The resulting transformations are
$(v,w)\mapsto(v,\sqrt{\frac{1+v^2+w^2}{2}})$ and
$(v,w)\mapsto\sqrt{\frac{2}{1+v^2+w^2}}\,(v,w)$, respectively.
Fig.~\ref{fig:distill0-3}c shows where this gives an advantage over the
one-copy protocol Fig.~\ref{fig:distill0-3}b. Obviously, two-copy
protocols do neither allow to distill separable state to $\rhoSep$ nor do
they allow to distill entangled states to $\VEPR$.

For three copies, though, the following two pairs of POVM operatos
provide a way to distill all states:
$$
M_A=M_B\propto\left(\begin{array}{cccccccc}
    0&1&1&0&1&0&0&0\\0&0&0&1&0&1&1&0
\end{array}\right)
$$
distills all separable states to $\rhoSep$ by virtue of 
$$
(v,w)\mapsto\left(v+\frac{v-v^3}{1+2v^2+2w^2},\frac{w(2+2v^2+w^2)}{1+2v^2+2w^2}\right)
$$
whereas
\begin{eqnarray*}
M_A'&\propto&\left(\begin{array}{cccccccc}
    0&1&1&0&-1&0&0&0\\0&0&0&1&0&1&1&0
\end{array}\right)\ ,
\\
M_B'&\propto&\left(\begin{array}{cccccccc}
    0&1&1&0&1&0&0&0\\0&0&0&-1&0&1&1&0
\end{array}\right)
\end{eqnarray*}
distills entangled states towards a $\VEPR$, and 
$$
(v,w)\mapsto
\left(\frac{v(6+3v^2-2w^2)}{3+6v^2+6w^2},\frac{w(6-2v^2+3w^2)}{3+6v^2+6w^2}\right)\ .
$$
This is illustrated in Fig.~\ref{fig:distill0-3}d.

%%%%%%%%%%%%%%%%%%%%%%%%%%%%%%%%%%%%%%%%%%%%%%%%%%%%%%%%%%%
%%%%%%%%%%%%%%%%%%%%%%%%%%%%%%%%%%%%%%%%%%%%%%%%%%%%%%%%%%%
\section{SiV as a resource
\label{section:resource}}
%%%%%%%%%%%%%%%%%%%%%%%%%%%%%%%%%%%%%%%%%%%%%%%%%%%%%%%%%%%
%%%%%%%%%%%%%%%%%%%%%%%%%%%%%%%%%%%%%%%%%%%%%%%%%%%%%%%%%%%

%**********************************************************
\subsection{Introduction}
%**********************************************************

In its standard form, i.e., as a singlet, EoE can be used to teleport
quantum bits and thus allows to overcome the 
restriction to LOCC. In this section, we will show that SiV
is a resource in very much the same way, namely it allows to overcome the
restrictions imposed by SSR in a bipartite setting. Despite the
similarities, there are some major differences. Firstly, while for EoE
there only exists one standard form (the maximally entangled state
depending on the dimension of the system), for SiV there exist two
different standard states: singlets $\ket{0}\ket N+\ket N\ket0$ with SiV
$N^2$ (as in Corollary~\ref{theorem:qubit-resources}) and the Gaussian
distributed states with large variance (as in
Theorem~\ref{theorem:EoE-SiV}).  Second, there are no pure states which
carry only SiV---SiV as a resource which is independent of EoE only exists
for mixed states where resources are difficult to quantify. Finally, we
will find that we need an infinite amount of nonlocality in order to
completely overcome the restrictions imposed by the SSR---this is
fundamentally different from EoE where one ebit of entanglement is
sufficient to perfectly teleport one quantum bit.

In order to demonstrate (and partly quantify) that SiV is useful to
overcome the restrictions imposed by the SSR, we will use the tasks of
distinguishing and teleportation. It has been shown that with SSR there
exists a perfect data hiding protocol~\cite{terhal:datahiding} which
allows to encode a classical bit in a bipartite state such that it cannot
be revealed by LOCC~\cite{verstraete:SSR-datahiding}.  This protocol can
be extended to a protocol hiding $\log N$ bits in the Fourier states
\begin{equation}
\label{eq:datahiding}
\ket{\zeta_k^N}=\frac{1}{\sqrt{N}}
    \sum_{n=0}^{N-1}e^{2\pi ikn/N}\ket{n}_A\ket{N-1-n}_B\ .
\end{equation}
These states can be distinguished perfectly by LOCC if no SSR are present
(therefore, both parties measure in the Fourier basis and compare their
outcomes), but with SSR, they become totally indistinguishable (see 
Theorem~\ref{theorem:indist} below). 

The second task we use to show that SiV is a resource is teleportation of
a state with nonconstant local particle number: Alice holds one half of a
state $\ket{\phi}_{AC}$ which she wants to teleport to Bob using an in
general mixed helper state $\rho_{AB}$. Clearly, one ebit of EoE is
necessary for this task, but if $V(\ket{\phi})>0$, also SiV is
needed~\cite{verstraete:SSR-datahiding}. We will show that one ebit of EoE
is still sufficient, but the amount of SiV has to grow superlinearly with
$V(\ket\phi)$ and is infinite for perfect teleportation.

%**********************************************************
\subsection{A general protocol}
%**********************************************************

Let us first quote a Theorem from~\cite{verstraete:SSR-datahiding} which
will be very useful in the following.
\begin{theorem}[\cite{verstraete:SSR-datahiding}]
\label{theorem:indist}
In the presence of superselection rules, the states $\rho$ and $\mathcal
N_{A|B}(\rho)$ cannot be distinguished by LOCC. Here, $\mathcal N_{A|B}$
is the ``dephasing map''
$$
\mathcal N_{A|B}(\rho)=
\sum_{n_A,n_B}P_{n_A}^A\otimes P_{n_B}^B\rho P_{n_A}^A\otimes P_{n_B}^B\ ,
$$
with $P_{n_X}^X$ the local projector onto the subspace with $n_X$
particles.
\end{theorem}
\vspace*{-0.01cm}

By this theorem, we can highly restrict the class of allowed protocols.
Let us show this for the task of distinguishing, where Alice and Bob
initially share $\rho=\ket{\zeta_k^N}\bra{\zeta_k^N}\otimes\sigma$
[cf.\ Eq.~(\ref{eq:datahiding})], and
they have to determine $k$. At the end of the protocol, Alice and Bob get
an answer $k'$ according to a probability distribution $\{p_{k'}\}$.  But
if they had started with $\mc N_{A|B}(\rho)$ instead,
Theorem~\ref{theorem:indist} tells us that the probability distribution of
their outcomes would have been just the same.  Therefore, Alice and Bob
can start their protocol by measuring their particle number operators
$\hat N_A$ and $\hat N_B$---if they discard their outcomes, they just
implemented $\mc N_{A|B}$, and their knowledge of $N_A$ and $N_B$ will not
affect the \emph{average} probability distribution which is solely
relevant unless the figure of merit is nonlinear.  The same holds for the
teleportation scenario with respect to the partition $A|BC$, i.e., in this
case only Alice is allowed to measure her particle number (this map is
actually weaker than $\mc N_{A|BC}$).  Note that for a pure state, 
measuring $\hat N_A$ also determines $N_{BC}$ (and thus implements $\mc
N_{A|BC})$, which is different in the mixed state case and closely
connected to the fact that mixed states alone are not sufficient for
teleportation.

%**********************************************************
\subsection{Pure states}
%**********************************************************

Assume Alice wants to teleport her share of the state
$$
\ket\phi=\sum_n\alpha_n\ket n_A\ket{-n}_C
$$
to Bob using 
$$
\ket\psi=\sum_m\beta_m\ket m_A\ket{-m}_B\ .
$$
Here, we use a simplified notation, where $-\infty<n,m<\infty$, 
$\sum|\alpha_n|^2=\sum|\beta_m|^2=1$, and the support of the $\alpha_n$,
$\beta_m$ is bounded below such that the particle number can be made
positive by adding ancillas. As shown before, Alice can start any protocol
by measuring $N_A=K$, yielding
\begin{equation}
\label{eq:pseudotelep-after-NA}
\sqrt{p_K}\ket{\chi_K}_{ABC}=\sum_{n+m=K}\alpha_n\beta_m\ket{n,m}_A\ket{-n}_C\ket{-m}_B
\end{equation}
with included probability $p_K$. If Alice now measures in the Fourier
basis and communicates her result, the originally tripartite state can be
reconstructed by Bob and Charlie; this strategy is optimal as no
information is lost and the state gets less nonlocal.
Up to shifts in the particle number, the state is then
$$
\sqrt{p_K}\ket{\chi_K}_{BC}=\sum_{n}\alpha_n\beta_{K-n}\ket{n}_B\ket{-n}\ .
$$
We will use the average entanglement fidelity as the figure of merit,
$$
\bar F=
    \left\langle\sum_Kp_K\left|\langle\phi\ket{\chi_K}\right|^2\right\rangle
    =\sum_{\Delta}\Pi(\Delta)C(\Delta)
$$
where $\Pi(\Delta)=\sum_n\left\langle p_np_{n+\Delta}\right\rangle$
($p_n=|\alpha_n|^2$) and $C(\Delta)=\sum_m\beta_m^*\beta_{m-\Delta}$.
The average $\langle\cdot\rangle$ is taken over all states $\phi$, where
for teleportation we assume a unitarily invariant distribution.
It is straightforward to check that local filtering operations cannot
increase $\bar F$. Also, for the task of distinguishing it can be shown
that $\bar F$ gives the optimal success probability for the inconclusive
case~\cite{chefles:overview}. For distinguishing,
$\Pi_D(\Delta)=\max(N-|\Delta|,0)/N^2$, while for teleportation,
$\Pi_T(\Delta)=[\max(N-|\Delta|,0)+\delta_{\Delta,0}]/N(N+1)$~%
~\cite{banaszek:haar-avg}. In both cases, $\alpha_n\ne0$ for 
$n=0,\dots,N-1$.

We will analyze two natural types of helper states:
states with constant distribution $\beta_m=1/\sqrt{M}$, $m=0,\dots,M-1$,
and states with Gaussian distribution with variance $V(\psi)$. One finds
$C_C(\Delta)=\max(M-|\Delta|,0)/M$ resp.\
$C_G(\Delta)=\exp[-\Delta^2/2V]\approx 1-\Delta^2/2V$. The resulting error
probabilities for all four cases are given in Table~\ref{table:p_err}.
Note that for the Gaussian distributed $\ket\psi$, in both cases
$$
p_\mathrm{err}=\frac{\big\langle V(\phi)\big\rangle}{4V(\psi)}
$$ 
holds, i.e., the error probability is given by the \emph{ratio} of the
variances. (Actually, this even holds without taking the
average over $\phi$.)

In all cases, the error vanishes only if the size of the
helper state grows superlinearly with the size of the unknown state; thus,
the scaling of SiV as a resource is unfavorable compared to the behaviour
of EoE. This is a direct consequence of the direct sum structure in~%
(\ref{eq:OABNBipSSR}) which is opposed to the tensor product structure
leading to EoE: while with a tensor product structure, $N$ particles
generate
a $2^N$-dimensional Hilbert space, for the direct sum structure the underlying
space only has dimension $N+1$. This also holds for mixed states,
where this is the size of the largest coherent subspace. For the same
reason, the data hiding scheme Eq.~(\ref{eq:datahiding}) is optimal in the
sense that the \emph{available} Hilbert space has only dimension $N$.

%#############################################
\begin{table}[t]
\begin{tabular}{l||l|l}
&\multicolumn{2}{c}{task}\\
    \cline{2-3}
\rule[-1ex]{-0cm}{3.7ex}
helper&\multicolumn{1}{c}{distinguish}&\multicolumn{1}{c}{teleport}\\
    \cline{2-3}\\[-2.1ex]\hline
constant&
    \rule[-1.8ex]{-0cm}{5.3ex}\hspace{1em}
	    $p_\mathrm{err}=\frac{(N+1)(N-1)}{3MN}$\hspace*{1em}&
    \hspace{1em}$p_\mathrm{err}=\frac{N}{3M}$\hspace*{1em}\\
    \hline
Gaussian&
    \rule[-1.8ex]{-0cm}{5.3ex}\hspace{1em}
    $p_\mathrm{err}=
	\frac{(N+1)(N-1)}{12V(\psi)}$\hspace*{1em}&
    \hspace{1em}$p_\mathrm{err}=
	\frac{N(N-1)}{12V(\psi)}$\hspace*{1em}
\end{tabular}
\caption{\label{table:p_err}
Error probabilities for distinguishing and teleportation where the helper
state is either a maximally entangled state with Schmidt number $N$ or a
Gaussian distributed state with variance $V(\psi)$.
}
\end{table}
%#############################################

\subsection{Mixed states}

Let us now demonstrate that separable mixed states with SiV can also be
used as a shared reference frame~\cite{verstraete:SSR-datahiding}.  First,
we demonstrate how Alice and Bob can use the state $\rhoSep$ to
distinguish the states $\ket{01}\pm\ket{10}$. By very much the same
argument as before Alice and Bob can start their protocol by measuring
their local particle number.  By \emph{adding} their outcomes, Alice and
Bob immediately know whether they are dealing with the
$\ket0\ket0$/$\ket1\ket1$ part of $\rho_\mathrm{sep}$ or with the $\VEPR$.
In the first case all information is lost while in the second case they
can just proceed as if they had started with $\VEPR$ itself. This case
occurs with probability $1/2$, i.e., all the SiV contained in
$\rho_\mathrm{sep}$ can be used. Clearly, this protocol can not be used
for teleportation as $\hat N_{BC}$ cannot be implemented locally.

Let us now show that separable but nonlocal states can be used to overcome
locality constraints arbitrarily well, i.e., they can serve as arbitrarily
precise reference frames. Therefore, we use the separable
state~\cite{verstraete:SSR-datahiding,rudoph:mixed-coherent}
$$
\rhoCoh(\alpha)=\int\frac{\mathrm{d}\phi}{2\pi}
    \ket{\alpha e^{i\phi}}\bra{\alpha e^{i\phi}}\otimes
    \ket{\alpha e^{i\phi}}\bra{\alpha e^{i\phi}}
$$
where for $\alpha>0$, 
$$
\ket{\alpha e^{i\phi}}=e^{-\alpha^2/2}\sum_{n=0}^\infty
    \frac{\alpha^n}{\sqrt{n!}}e^{in\phi}
$$
is a coherent state with amplitude $\alpha e^{i\phi}$.  It has been
shown~\cite{verstraete:SSR-datahiding} that for $\alpha\rightarrow\infty$,
$\rhoCoh(\alpha)$ can be used to distinguish $\ket{01}\pm\ket{01}$ with
arbitrary precision. In the following, we will show that this state
together with one $\CEPR$ can be used to perfectly teleport a state with
nonconstant local particle number and therefore may serve as an
arbitrarily precise reference frame.

First, let us use Theorem~\ref{theorem:VF-additive} to
show that this state has indeed infinite SiV for
$\alpha\rightarrow\infty$. Therefore, it is enough to note that
\begin{eqnarray*}
\rhoCoh(\alpha)&=&\sum_{N=0}^{\infty}p_N\ket{\theta_N}\bra{\theta_N}\ ;\quad
p_N=e^{-2\alpha^2}\frac{(2\alpha^2)^N}{N!}\\
&&\ket{\theta_N}=
    \frac{1}{\sqrt{2^N}}\sum_{n=0}^N\left(N\atop n\right)^{1/2}
    \ket{n,N-n}\ ,
\end{eqnarray*}
and thus
$V_F^\SSR(\rhoCoh(\alpha))=V_c^\SSR(\rhoCoh(\alpha))=\alpha^2/2\rightarrow\infty$
for $\alpha\rightarrow\infty$. In is interesting to note that each of the
$\ket{\theta_N}$ approximates a state with Gaussian distribution such that
$\rhoCoh(\alpha)$ might be considered as the EoE-free mixed state version
of Gaussian distributed states.

In order to see how a mixed state can be used to teleport, let Alice and
Charlie initially share $\ket\phi=\alpha\ket{01}+\beta\ket{10}$ (the proof
is completely analogous for qu-$d$-its), and
assume Alice wants to teleport her
share to Bob. Therefore, Alice and Bob are provided with an $\CEPR_{AB}$ and
with some mixed state
\begin{equation}
\label{eq:generic-refframe}
\rho=\sum_{{n,m,n',m'}\atop{n+m=n'+m'}}\rho_{n,m}^{n',m'}
	\ket n_A\bra{n'}\otimes\ket m_B\bra{m'}\ ,
\end{equation}
where the condition $n+m=n'+m'$ comes from the SSR,
Eq.~(\ref{eq:OisPnOPn}).  For simplicity, let us assume that all
$\rho_{n,m}^{n',m'}$ are nonnegative.  Alice once more starts by measuring
her local particle number operator on $\ket\phi\bra\phi\otimes\rho$.  (In
this step, we do not have to care about the $\CEPR$ which has constant
local particle number.) For a measurement outcome $n$, the resulting state
(probability included) is 
\begin{eqnarray*}
\sum_{m}\hspace*{-1em}&&\Big[
    |\alpha|^2\rho_{n,m-1}^{n,m-1}\ket{0,n}_A\bra{0,n}\otimes\ket{m-1}_B\bra{m-1}\otimes\ket{1}_C\bra{1}
    \\
    &&+|\beta|^2\rho_{n-1,m}^{n-1,m}\ket{1,n-1}_A\bra{1,n-1}\otimes\ket{m}_B\bra{m}\otimes\ket{0}_C\bra{0}
    \\
    &&+\alpha\beta^*\rho_{n,m-1}^{n-1,m}\ket{0,n}_A\bra{1,n-1}\otimes\ket{m-1}_B\bra{m}\otimes\ket{1}_C\bra{0}
    \\
    &&+\alpha^*\beta\rho_{n-1,m}^{n,m-1}\ket{1,n-1}_A\bra{0,n}\otimes\ket{m}_B\bra{m-1}\otimes\ket{0}_C\bra{1}\Big]
\end{eqnarray*}
As Alice's share now has constant particle number and lies within a
two-dimensional subspace, she can use the $\CEPR_{AB}$ to teleport her share to
Bob. If we label the two teleported basis states $\ket{\hat a}=\ket{0,n}$,
$\ket{\hat b}=\ket{1,n-1}$, Bob and Charlie then share the state
\begin{eqnarray*}
\sum_{m}\hspace*{-1em}
    &&\Big[|\alpha|^2\rho_{n,m-1}^{n,m-1}\ket{\hat a,m-1}_B\bra{\hat a,m-1}\otimes\ket{1}_C\bra{1}
\\
&&    +|\beta|^2\rho_{n-1,m}^{n-1,m}\ket{\hat b,m}_B\bra{\hat b,m}\otimes\ket{0}_C\bra{0}
\\
&&    +\alpha\beta^*\rho_{n,m-1}^{n-1,m}\ket{\hat a,m-1}_B\bra{\hat b,m}\otimes\ket{1}_C\bra{0}
\\
&&    +\alpha^*\beta\rho_{n-1,m}^{n,m-1}\ket{\hat b,m}_B\bra{\hat a,m-1}\otimes\ket{0}_C\bra{1}\Big]
\end{eqnarray*}
Bob now projects onto the subspaces spanned by the pairs of states
$\ket{0_m}\equiv\ket{\hat a,m-1}$ and $\ket{1_m}\equiv\ket{\hat b,m}$ and obtains
\begin{eqnarray}
&&|\alpha|^2\rho_{n,m-1}^{n,m-1}\ket{0}_B\bra{0}\otimes\ket{1}_C\bra{1}
\nonumber \\
    &+&|\beta|^2\rho_{n-1,m}^{n-1,m}\ket{1}_B\bra{1}\otimes\ket{0}_C\bra{0}
\nonumber\\
    &+&\alpha\beta^*\rho_{n,m-1}^{n-1,m}\ket{0}_B\bra{1}\otimes\ket{1}_C\bra{0}
\nonumber\\
    &+&\alpha^*\beta\rho_{n-1,m}^{n,m-1}\ket{1}_B\bra{0}\otimes\ket{0}_C\bra{1}
\label{eq:after-teleportation}
\end{eqnarray}
(where we omitted the subscript $m$).
By looking at the average fidelity with the original state, we find that
the error vanishes iff 
\begin{equation}
\sum_{n,m}\rho_{n,m-1}^{n,m-1}=
\sum_{n,m}\rho_{n-1,m}^{n-1,m}=\sum_{n,m}\rho_{n-1,m}^{n,m-1}\ .
\label{eq:fid1-teleport}
\end{equation}
Since
$\rho$ is positive this implies that 
$\rho_{n,m-1}^{n,m-1}\approx\rho_{n-1,m}^{n-1,m}\approx\rho_{n-1,m}^{n,m-1}$
for most $n,m$, as one would expect from
Eq.~(\ref{eq:after-teleportation}). It is straightforward to check that 
Eq.~(\ref{eq:fid1-teleport}) holds for $\rhoCoh(\alpha)$ for
$\alpha\rightarrow\infty$, and that the $2\times2$ subblocks of the density matrix really
approximate pure states.

One might expect that $N\rightarrow\infty$ copies of $\rhoSep$ could be
used just the same way, but the situation is quite different: filtering
operations which bring $\rhoSep^{\otimes N}$ into a
form~(\ref{eq:generic-refframe}) destroy the off-diagonal elements of the
density matrix with high probability so that~(\ref{eq:fid1-teleport})
cannot be satisfied; therefore it is questionable whether
multiple copies of $\rhoSep$ can be used as an arbitrarily precise
reference frame. On the other hand, this is not so much different from the
pure state scenario: while multiple copies of a $\VEPR$ might indeed be
used as a perfect reference frame, these states carry an amount of
entanglement which grows \emph{linearly} with the precision of the reference
frame, whereas a single Gaussian distributed state with large SiV only has
\emph{logarithmic}---and
thus in some sense vanishing---entanglement and is therefore much closer
to the case of separable reference frames.

Let us note that the teleportation scenario can be altered by joining $B$
and $C$. This is no longer teleportation, of course, and can be
accomplished by LOCC without SSR. On the other hand, it is still an
impossible task if SSR are present and is thus suitable to characterize mixed
states as reference frames without the need for additional entanglement.

\subsection{Hiding quantum states}

Let us close by showing that the data hiding
protocol given in~\cite{verstraete:SSR-datahiding} resp.\ its extension
Eq.~(\ref{eq:datahiding}) can be used to construct a mixed state scheme to
hide quantum data as well. At a first glance, one might try to encode the
two degrees of freedom of a qubit in the phases of the state
$\ket{02}+e^{i\phi}\ket{11}+e^{i\phi'}\ket{20}$, but this cannot be
accomplished by a linear map. Therefore, we encode the qubit
$\alpha\ket{01}+\beta\ket{10}$ in one of the states
$
\ket{\phi_0}=\alpha\ket{01}+\beta\ket{10}$, $
    \ket{\phi_1}=\beta\ket{01}+\alpha\ket{10}
$
with equal probabilities which is then distributed between Alice and Bob.
Additionally, Alice and Bob are provided with a state
$\ket{\psi_{0/1}}=\ket{02}\pm\ket{20}$ which encodes the state Alice and
Bob actually share. Thus, Alice and Bob share a state which they cannot
distinguish from the totally mixed state by LOCC
(Theorem~\ref{theorem:indist}), but they can perfectly recover the
original qubit if they join.  This scheme can be extended to hide
$N$-level quantum states using one of the states
$$
\ket{\phi_k}=\sum_{n=0}^{N-1}\alpha_{n+k\,\mathrm{mod}\,N}\ket{n,N-1-n}\ ;
    \quad k=0,\dots,N-1\ .
$$
Together with the state encoding $k$, $N^2-1$ particles are needed, and
the associated Hilbert space dimension is $N^2$.

\section{Conclusions}

Adding restrictions to the operations permissible on a quantum system
gives rise to a new resource which in turn allows to overcome this
restriction. The restriction to LOCC, for example, leads to EoE as a
nonlocal resource. Adding SSR to a bipartite system leads to an additional
resource, the superselection induced variance SiV. We could show that SiV
and EoE together completely characterize
bipartite states in the asymptotic limit. Thereby, two different
kind of standard forms arise, namely singlets and
Gaussian distributed states with logarithmic EoE. 

The search for states which only carry SiV led us to mixed states, where we
considered entanglement and variance of formation. We could show that the
concept of entanglement does not have to be changed and thus there exist
states which carry SiV but no EoE, and we provided explicit formulas for
the case of qubits. As to distillation, we could show that both EoE and
SiV can be distilled, and we provided various ways to do that. Thereby, we
found that there exist mixed standard states for SiV which do not carry
EoE.  While it is possible to extend recurrence
protocols such that they work with SSR by using a third copy as a
reference frame, it is unlikely that protocols with asymptotic yield work.

Finally, we showed that SiV is a resource which allows to overcome the
restrictions imposed by the SSR, but we also saw that there are
fundamental differences to EoE as the size of the reference frame has to
grow superlinearly with the problem size, which is due to direct sum
structure of the underlying Hilbert space.

\section*{Acknowledgments}

We acknowledge helpful discussions with K.~G.~Vollbrecht and M.~M.~Wolf.
This work was supported in part by the E.C.\ (RESQ QUIPRODIS) and
the Kom\-pe\-tenz\-netz\-werk ``Quanteninformationsverarbeitung'' der
Bay\-er\-ischen Staats\-re\-gie\-rung.

\appendix*

\onecolumngrid

\section{Proof of Eq.~(\ref{eq:P_n-log-creation})}

In order to show Eq.~(\ref{eq:P_n-log-creation}), we need the inequality
$S(p_i\rho_i)\le p_iS(\rho_i)+H(p_i)$ (see, e.g.,~\cite{nielsen-chuang}
for a proof). Furthermore, note that
$P_n=\sum_{i=0}^nP_i^A\otimes P_{n-i}^B$. Toghether, this gives the
estimate

\begin{eqnarray*}
E\left(\frac{P_n\ket\psi}{\sqrt{\bra\psi P_n\ket\psi}}\right)
    &=&
S\left(\frac
    {\tr_B\sum_{i=0}^n P_i^A\otimes P_{n-i}^B\ket\psi\bra\psi P_i^A\otimes P_{n-i}^B}
    {\bra\psi P_n\ket\psi}
    \right)
\\
&\le&
    \sum_{i=0}^n
    \frac{\bra\psi P_i^A\otimes P_{n-i}^B\ket\psi}{\bra\psi P_n\ket\psi}
    \ S\left(\frac
    {\tr_B P_i^A\otimes P_{n-i}^B\ket\psi\bra\psi P_i^A\otimes P_{n-i}^B}
    {\bra\psi P_i^A\otimes P_{n-i}^B\ket\psi}
    \right)
    +H\left(\left\{
	\frac{\bra\psi P_i^A\otimes P_{n-i}^B\ket\psi}
	{\bra\psi P_n\ket\psi}
    \right\}_{i=0}^N\right)\ .
\end{eqnarray*}
Clearly, the Shannon entropy $H$ is bounded by $\log(n+1)\le\log(N+1)$, and
thus the l.h.s.\ of Eq.~(\ref{eq:P_n-log-creation}), i.e., the
entanglement averaged over $n$, is bounded by
$$
\sum_{n=0}^N
\sum_{i=0}^n
    \bra\psi P_i^A\otimes P_{n-i}^B\ket\psi
    \ S\left(\frac
    {\tr_B P_i^A\otimes P_{n-i}^B\ket\psi\bra\psi P_i^A\otimes P_{n-i}^B}
    {\bra\psi P_i^A\otimes P_{n-i}^B\ket\psi}
    \right)
    +\log[N+1]\ .
$$

The sum can be extended to $i=0,\dots,N$, $n-i=0,\dots,N$ as $\ket\psi$
has at most $N$ particles, and by the convexity of the von Neumann
entropy, Eq.~(\ref{eq:P_n-log-creation}) follows.

\twocolumngrid


\begin{thebibliography}{30}
\bibitem{footnote1}
E.g, for class of $2\times2$ qubit states given by $
  \ket\phi=\alpha\ket{00}_A\ket{11}_B+ \beta\ket{01}_A\ket{10}_B+
  \beta\ket{10}_A\ket{01}_B+ \alpha\ket{11}_A\ket{00}_B $ with
  $\alpha=\sqrt{p/2}$, $\beta=\sqrt{(1-p)/2}$, and $p<1/2$, $V_F^\SSR$ is not
  superadditive with respect to the two subsystems, since $V(\ket\phi)=4p$ and
  $V_F^\SSR(\Tr_i[\ket\phi\bra\phi])=4p(1-p)$.
\bibitem{footnote:stdform-special}
Some special cases for $\rho$ have to be considered
  separately. Note that if $v=0$ $\rho$ can be created by LOCC, and $v=0$
  follows if $w_{01}=0$ or $w_{10}=0$. If $w_{00}=w_{11}=0$, filtering still
  works in both directions if $w_{00}$ and $w_{11}$ are deleted from the
  filtering operators. In case $w_{00}=0$, $w_{11}>0$ or vice versa, the
  situation gets more complicated. In order to distill $\rho$, one adds
  $\ket{00}\bra{00}$ with some weight which does not destroy the entanglement;
  then, $\rho$ can be distilled to $\VEPR$, from where it can be easily
  reconstructed.
\end{thebibliography}
\end{document}